\documentclass[pre,a4paper,twocolumn,superscriptaddress,%
floatfix]{revtex4}

\usepackage[T1]{fontenc}
\usepackage[utf8]{inputenc}

\usepackage{graphicx}
\usepackage{color}
\usepackage{amsmath}

\newcommand{\tac}{\ensuremath{T_{\mathrm{ac}}}}
\newcommand{\dtac}{\ensuremath{A_T}}

\begin{document}

\title{Understanding temperature modulated calorimetry through studies of a
  model system}

\author{Jean-Luc Garden}
\affiliation{Univ. Grenoble Alpes, CNRS, Grenoble INP, Institut N\'{E}EL,
38000 Grenoble, France}

\author{Michel Peyrard}
\affiliation{Ecole Normale Supérieure de Lyon (ENSL), CNRS, Laboratoire de
  Physique, F-69342 Lyon, France}

\date{\today}

\begin{abstract}
Temperature Modulated calorimetry is widely used but still raises some
fundamental questions. In this paper we study a model system as a test sample
to address some of them. The model has a nontrivial spectrum of relaxation
times. We investigate temperature modulated calorimetry at constant average
temperature to precise the meaning of the frequency-dependent heat capacity,
its relation with entropy production, and how such measurements can observe
the aging of a glassy sample leading to a time-dependent heat capacity.  The
study of the Kovacs effect for an out-of-equilibrium system shows how
temperature modulated calorimetry could contribute to the understanding of
this memory effect. Then we compare measurements of standard scanning
calorimetry and temperature-modulated calorimetry and show how the two
methods are complementary because they do not observe the same features. While
it can probe the time scales of energy transfers in a system, even in the
limit of low frequency temperature modulated calorimetry does not probe some
relaxation phenomena which can be measured by scanning calorimetry, as
suggested by experiments with glasses.

\end{abstract}

\maketitle

\section{Introduction}
\label{sec:intro}
Many calorimetry studies rely on a modulated heating power to determine
a frequency-dependent heat capacity of a sample
(that we henceforth call ``dynamic heat capacity'') \cite{KRAFTMAKHER}.
There are technical reasons such as the use of a
high modulation frequency so that the heat loss of the sample becomes
negligible and the high accuracy which can be reached in measurements which
use lock-in amplifiers and filters.
There are also more fundamental motivations. Modulated
techniques can be used to select among various timescales in the evolution
of the system, or to 
determine the heat capacity while the average
temperature of the sample is kept fixed and only very small oscillations are
imposed. It is also an interesting method to follow the time dependence of the
thermal properties of a system out of equilibrium, such as
measurements on glasses. Large improvements in the experimental methods have
been achieved \cite{GMELIN,HATTA} while theoretical analysis
  generated a lot of attention and controversy as attested by the special
  issue of the Journal of Thermal Analysis devoted to temperature
  modulated calorimetry in 1998 \cite{MENCZEL}.
Since then these methods are still raising
some fundamental questions \cite{GARDEN-REVIEW} to understand the meaning of
the measurements \cite{SCHAWE,READING1997}
which are sometimes described by a complex specific heat.
The puzzle becomes even
more complicated when relaxation phenomena in the sample are mixed with its
response to a modulated heat signal \cite{ANDROSCH,TOMBARI2007}.

\smallskip
Theoretical analyses from irreversible thermodynamics \cite{GARDEN-REVIEW} or
linear response theory \cite{NIELSEN} bring useful insights, but these
approaches could be usefully completed by the investigation of a system which
can be fully characterized and controlled. One feature which characterizes
physics is that it has often made progress by studying simple ``model''
systems, whether they are real systems such as the hydrogen atom in the
development of quantum mechanics or theoretical models such as the Ising model
for statistical physics.
At a first glance they may appear as too simple, but, because they allowed
physicists to identify the basic mechanisms behind the observations, they
turned out to provide elements for a basic understanding. Moreover some of the
simplest theoretical models such as the Ising model can even be applied to
quantitatively describe a large variety of real systems to a good
approximation. In this paper we show how a simple thermodynamic model can
clarify many questions which arise in the analysis of modulated calorimetry
experiments.

\smallskip
To be useful such a model should be sufficiently simple to allow a complete
analysis, but nevertheless rich enough to capture subtle effects which appear
in real experimental situations.
The idea to investigate the properties of a model system to clarify the
meaning of the dynamic heat capacity has already been explored in the study of
a bead-spring chain viewed as a model for a glass former
\cite{BROWN-JR2009,BROWN-JR2011}
however this system is too complex to allow a full analytical analysis.
Recently we showed that a three-state system
can play the role of the ``simplest complex system'' able to describe a large
variety of properties of glassy systems, for instance the subtle Kovacs effect
which demonstrates that thermodynamic variables are not sufficient to
characterize the state of some out-of-equilibrium systems such as glasses,
while allowing an analytical analysis of most of the phenomena \cite{PG}. In
this paper we show that this model can also provide a useful test system to
investigate the different contributions which enter in the signal measured in
a temperature modulated calorimetry experiment and to relate them to the
underlying physical phenomena. However we want to stress that,
  in the context of this paper, 
  the model is only a convenient tool for our analysis because it describes a
  system in which energy transfers occur at various time scales. This is the
  class of phenomena which are typically studied by temperature-modulated
  calorimetry. The possibility to derive analytical expressions for the
  response of this system, even when it is very far from equilibrium
  and shows strong relaxations, clarifies the origin of the different
  contributions which can be detected with temperature-modulated
  calorimetry and should hopefully help in the analysis of experimental data
  for various systems.

\smallskip
Section ~\ref{sec:model} introduces the three-state model which is at the
basis of this study and discusses its application to numerical simulations of
temperature modulated calorimetry experiments. Section \ref{sec:constantt}
analyzes temperature modulated 
calorimetry experiments carried at constant average temperature, starting from
the determination of the dynamic specific heat for an equilibrium state and then
studying out-of-equilibrium states reached after a temperature jump. We show
that a full analytical treatment is possible in these cases so that the
different contributions to the signal recorded in the measurements can be
precisely assigned. This brings further light on the meaning of
the frequency-dependent specific heat beyond
the linear response theory\cite{NIELSEN}
and the origin of the influence of temperature modulation on the properties
of some systems,  which has been observed experimentally \cite{JOHARI1999}.
Section \ref{sec:tmdsc} considers experiments in which a modulation is
superimposed to a temperature ramp, as in Modulated Temperature
Scanning Calorimetry (MTSC). Experiments detect a qualitative
difference between the specific heat recorded in a standard Differential
Scanning Calorimetry (DSC) experiment, which uses the ramp alone, and the
frequency-dependent specific heat obtained by MTSC \cite{LAARRAJ}. They can
be understood from simulations and analytical calculations using
the three-state model. Section
\ref{sec:discussion} is a concluding discussion of the meaning of our results
and their interest to analyze the data of temperature modulated calorimetry on
real systems.

\section{The model and its application in calorimetry}
\label{sec:model}

\begin{figure}[h]
  \centering
  \includegraphics[width=4cm]{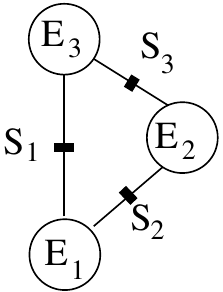}
  \caption{The three-state model. The circles schematize the metastable states
    with energies $E_i$ and the thick lines the barriers that separate them,
    with energies $S_i$.}
  \label{fig:model3st}
\end{figure}

The thermodynamic properties of complex systems can be approximately described
by models which focus on the minima of their free energy landscape
\cite{STILLINGER-82}. If this picture is completed by the values of the
barriers between the minima, dynamical properties can also be
investigated. Pushing these idea to the extreme leads to the
  two-level 
system, which is able to describe some surprising features of glasses. For
instance, in a glass which has been cooled very fast, some degrees of freedom
are trapped in high energy metastable states due to kinetic constraints. Then,
upon heating those states may relax so that, in a first stage, the energy
decreases while temperature increases, which is detected as a negative
heat capacity in the measurements \cite{BISQUERT}. The two-level system is
able to describe this phenomenon, observed for instance in $\mathrm{B_2O_3}$
\cite{DEBOLT}. However such an oversimplified system is not able to describe
more subtle properties of glasses, such as the Kovacs effect which
demonstrates that an out-of-equilibrium system is not fully characterized by
the knowledge of its thermodynamic variables \cite{PG}. Adding one metastable
state to get the three-state system shown on Fig.~\ref{fig:model3st}
is enough to describe such a
phenomenon. Moreover the three-state system is interesting in the context of
temperature modulated calorimetry because the spectrum of its relaxation times
has two 
characteristic times instead of the single relaxation time of the two-level
system. Therefore its dynamics is richer and can exhibit non trivial
time-dependencies beyond a simple exponential relaxation. The Kovacs effects is
only an example of such a situation \cite{PG}.

Of course a model which only considers the minima of the free energy landscape
cannot be complete. It only describes the {\em configurational} heat capacity
because it ignores other contributions to the energy, such as the vibrational
or electronic contributions. However, for glasses, the configurational heat
capacity is strongly dominant, as shown for instance by measurements on
poly(vynil acetate) (PVAC) \cite{TOMBARI2007}.

\smallskip

Let us denote by $E_i$ ($i = 1,2,3$) the energies of the three metastable
states.  The probabilities $P_i$ that the states $i$ are occupied are the
variables which define the state of the system. However, due to the constraint
$P_1 + P_2 + P_3 = 1$, the state is actually defined by two parameters
only. The two variables $P_1$ and $P_2$ are sufficient to characterize a
state of the system.

The equilibrium properties of the model are readily obtained from the
Gibbs canonical distribution. Its partition function is 
\begin{equation}
  \label{eq:z}
  Z = \sum_i e^{-E_i/T} \; ,
\end{equation}
if we measure the temperature $T$ in energy units (which is equivalent to
setting the Boltzmann constant to $k_B = 1$). When the system is in
equilibrium the occupation probabilities of the three states are 
\begin{equation}
  \label{eq:pieq}
  P_i^{\mathrm{eq}} = \frac{1}{Z} e^{-E_i/T} \; ,
\end{equation}
the average energy of the system is
\begin{equation}
  \label{eq:emoyeq}
  E^{\mathrm{eq}}(T) =
  \langle E(T) \rangle^{\mathrm{eq}} = \frac{1}{Z} \; \sum_i E_i e^{-E_i/T} \; ,
\end{equation}
and its equlibrium heat capacity is
\begin{equation}
  \label{eq:ceq}
  C^{\mathrm{eq}}(T) = \frac{d E^{\mathrm{eq}}(T)}{d T} =
  \frac{1}{T^2} \left[ \langle E^2(T) \rangle^{\mathrm{eq}}
    - \Big(\langle E(T) \rangle^{\mathrm{eq}}\Big)^2\right]
  \; ,
\end{equation}
where $\langle \; \rangle^{\mathrm{eq}} $ designates averages computed with
the equilibrium probabilities (Eq.~(\ref{eq:pieq})\,).
Viewing
this model as a simplified picture of the free energy landscape of a glass, we
assume that the transitions from one basin of attraction to another are
thermally 
activated over saddle points having energies $S_1$, for the transition between
$E_1$ and $E_3$, $S_2$ for the transition between $E_1$ and $E_2$, and $S_3$
for the transition between $E_2$ and $E_3$. 
Therefore the transition
probabilities are determined by a  set barriers $B_{ij}$,
for instance 
$B_{13} = S_1 - E_1$, $B_{31} = S_1 - E_3$, $B_{12} = S_2 - E_1$, and so on.
The energies of the saddle points
are assumed to be higher than the energies of the states that they separate so
that $B_{ij} >0$ for all $i,j$ pairs.

The rates of the thermally activated transitions are
\begin{equation}
  \label{eq:wij}
  W_{i \to j} = \omega_{ij} \; e^{- B_{ij}/T}
\end{equation}
where $\omega_{ij}$ are model parameters which have the dimension of inverse
time. As a result the thermodynamics of the model is expressed by equations
for the time-dependence of the occupation probabilities, which are of the form
\begin{align}
  \label{eq:dp1dt}
  \frac{d P_1}{dt} = &- P_1 \, \omega_{13} \, e^{-(S_1 - E_1)/T} +
                       P_3 \, \omega_{31} \, e^{-(S_1 - E_3)/T}  \nonumber \\
                     & - P_1 \, \omega_{12} \, e^{-(S_2 - E_1)/T} +
                  P_2 \, \omega_{21} \, e^{-(S_2 - E_2)/T} \; ,
\end{align}
and similar equations for $P_2$ and $P_3$.
The detailed balance conditions $\omega_{ij} = \omega_{ji}$ ensure the
existence of the equilibrium 
state. Detailed balance does not require $\omega_{13} = \omega_{12}$, however
we shall henceforth assume
\begin{equation}
  \label{eq:omeg1}
  \omega_{ij} = 1 \qquad \forall i, j (i \not= j) \;.
\end{equation}
Setting the common value of $\omega_{ij}$ to $1$ defines the time unit (t.u.)
for the system.

\bigskip
In a typical temperature modulated calorimetry experiment a physical system of
interest, the 
sample, is linked by a heat exchange coefficient $K$ to a thermal bath at
temperature $T_0$ (which could be time dependent if a thermal ramp is used). An
oscillatory power is transmitted to the sample, causing its temperature to
oscillate. Temperature and energy flow exchanged by the sample and its
environment are
monitored so that the variation of its temperature and energy versus time are
known, allowing a determination of its heat capacity. Our simulations use a
simpler scheme, which is adapted for a theoretical investigation. We {\em
  impose} the temperature $T(t)$ of the sample as a function
of time. Given an initial state,
this allows us to solve the equations for the time dependence of the
occupation probabilities of the three states with a
fourth order Runge-Kutta scheme, so that the energy $E(t)$ is
determined.

\smallskip
While the goal of calorimetry is to relate the variations of the energy and
temperature in a physical system, in temperature modulated calorimetry
experiments the energy is generally {\em controlled} through the modulated power
applied to the
sample, and temperature is {\em measured}. In our case the situation is
reversed because temperature is {\em controlled} and energy is {\em measured}
(or rather calculated). In both cases the heat capacity can be deduced from
the relation between energy and temperature but our simulations cannot claim
to address all the phenomena which take place in an actual calorimetry
experiment. In a complex system, including the three-state system, knowing the
energy does not fully determine the state of the system because several
configurations can lead to the same energy. One question which could be asked
is how does the energy split between the microscopic degrees of
freedom. Temperature Modulated calorimetry can try to answer this question by
measuring the 
response to different modulation frequencies, which is determined by the
physical processes which take place in the system and govern the energy
exchange between microscopic states. In the three-state system
the energy exchanges are imposed by the variations of temperature
that we impose and by the rules that define the transition probabilities
between states. Nevertheless, as far as the calorimetric measurements are
concerned, our studies can probe how the calorimetric signal is affected,
given the different time scales determined by the temperature of the system.

\section{Temperature Modulated calorimetry at constant average temperature}
\label{sec:constantt}

In this section we consider a sample, modeled by the three-state system, with
a modulated temperature
\begin{equation}
  \label{eq:Tmodul}
 T(t) = T_0 + \tac (t) \quad{\mathrm{with~~}}  \tac(t) = \dtac \sin \omega t
\end{equation}
where $T_0$ is a
constant. This may occur in
various experimental situations. (i) If the system is in equilibrium at
temperature $T_0$, adding a small modulation ($\dtac \ll
T_0$) allows the measurement of its response at different frequencies to probe
the spectrum of the thermal relaxations of the sample. (ii) If the system is
strongly out-of-equilibrium at the start of the experiment, one may be
interested in its thermal aging, i.e.\ the time evolution of its heat capacity
$C(T_0,t)$. Some temperature change is necessary to measure the thermal
response. Choosing a temperature modulation $\tac(t)$ has a double
interest. First the average temperature of the sample is not modified so that,
if $\dtac \ll T_0$ one measures the specific heat at
$T_0$ to a good accuracy, and second, as
the measurement depends on $\omega$, the time dependence of the dominant
relaxation phenomena within the sample can be followed. (iii) If $\dtac$
is large, nonlinearities are excited and we shall show that
this can lead to some relaxation {\em even if the system starts from an
  equilibrium state}, as observed in some experimental investigations
\cite{WANG-JOHARI}.

\subsection{Spectrum of the fluctuations of energy transfers in an equilibrium system}
\label{se:spectrumequil}

Using the three-state system as a ``sample'' we can simulate a temperature
modulated 
calorimetry experiment that probes the time scales at which the energy can be
transferred between the degrees of freedom of a physical system.

In these calculations we used the following parameters: The energies of the
three metastable states are $E_1 = -0.40$, $E_2 = -0.25$, $E_3 = 0$.
The energies of the saddle points are $S_1 = 0.40$ between states 1 and 3,
$S_2 = 0.30$ between states 1 and 2, $S_3 = 0.25$ between states 2 and 3
(Fig.~\ref{fig:model3st}). As the model has not been tailored to any particular
physical system, the energy scale is irrelevant and the values have been
chosen arbitrarily. Temperatures are measured in energy units.
Other values would of course quantitatively change the results, but, as long
as the barriers $S_j - E_i$ for all the 
possible transitions are positive and the ratios of the investigated
temperatures and energies stay in the same range, the main features of the
results would not be affected.

\begin{figure}[h]
  \centering
  \includegraphics[width=7cm]{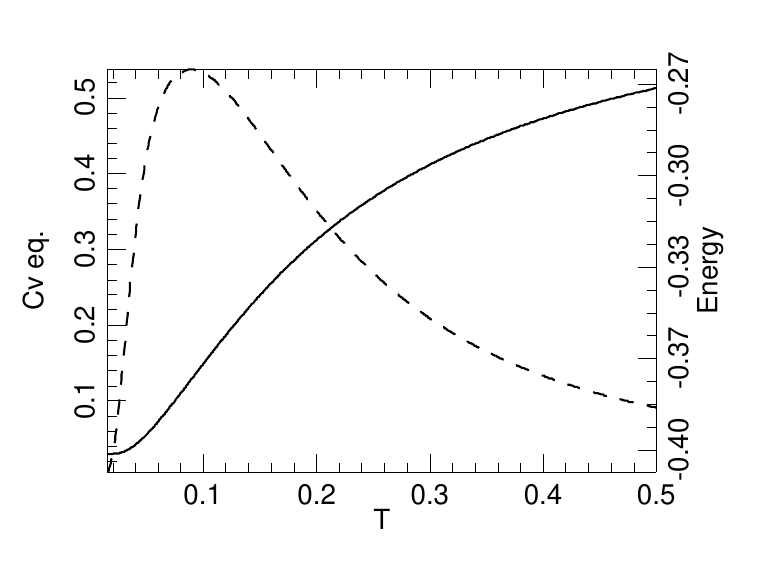}
  \caption{Equilibrium heat capacity (dashed line) and equilibrium
    energy (full line)
    versus temperature (in energy units) for the three-state model with
  $E_1 = -0.40$, $E_2 = -0.25$, $E_3 = 0$ and $S_1 = 0.40$, $S_2 = 0.30$, $S_3
  = 0.25$.  }
  \label{fig:ceequil}
\end{figure}

Figure \ref{fig:ceequil} shows the equilibrium heat capacity and
the equilibrium energy of the model
versus temperature for this parameter set. A temperature modulated calorimetry
measurement 
is simulated by imposing a variation of $T(t)$ according to
Eq.~(\ref{eq:Tmodul}) and solving the equation (\ref{eq:dp1dt}) and the similar
equation for $P_2$ (and $P_3 = 1 - P_1 - P_2$ is then obtained too)
with a 4th order Runge-Kutta
algorithm, using the time step $\delta t = 0.01$.
We have chosen $T_0 = 0.125$ and, in this section, the initial state is the
equilibrium state at $T=T_0$. Therefore, at this stage, no
  aging phenomena are involved.

\begin{figure}[h]
  \centering
  \includegraphics[width=7cm]{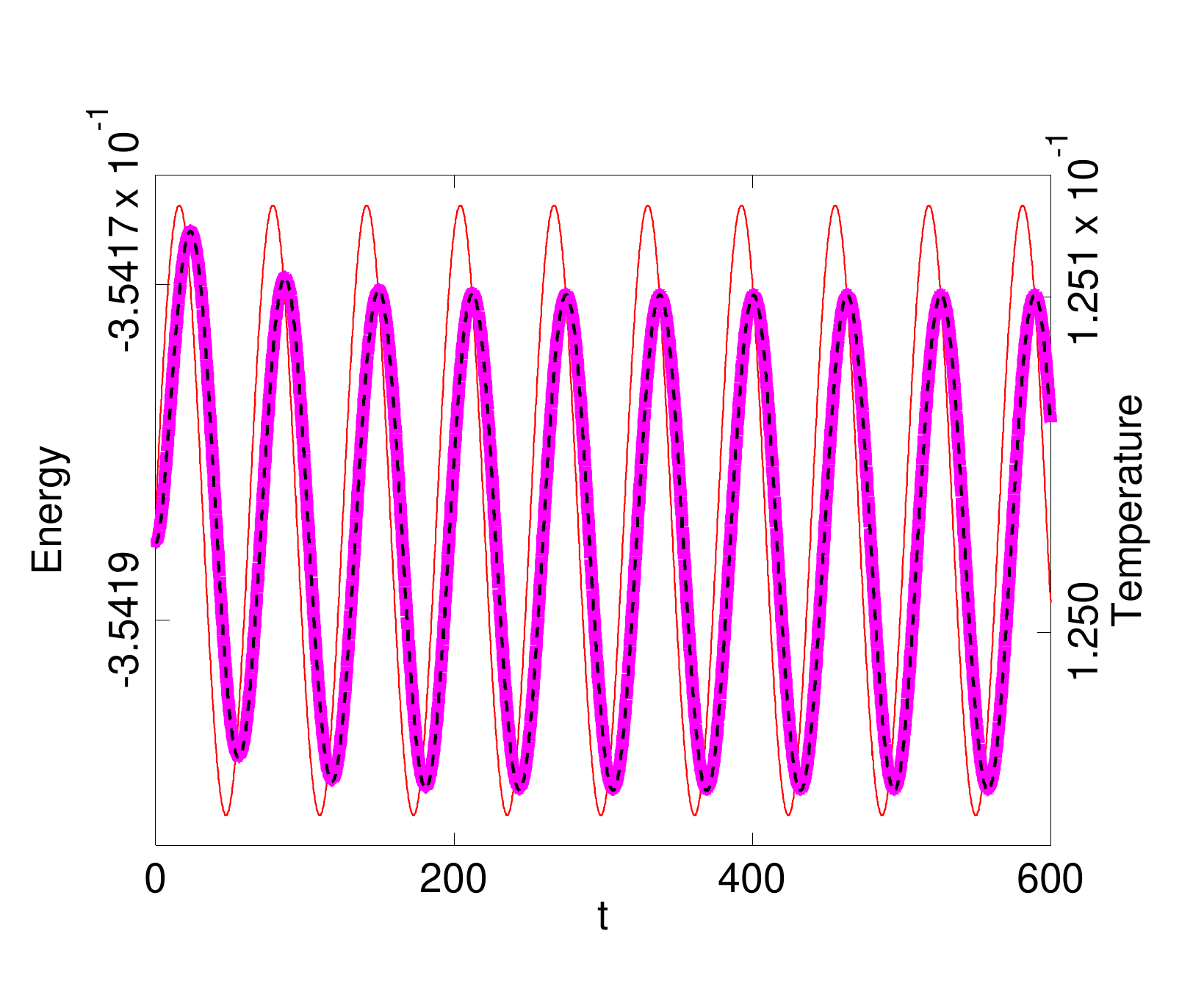}
  \caption{Energy and temperature versus time recorded in a simulation of the
    three state system, starting from an equilibrium state at
    temperature $T_0 = 0.125$, subjected to a
    modulated temperature according to Eq.~(\ref{eq:Tmodul}) with
    $\omega = 0.1$, $\dtac = 10^{-4}$. The thin red curve shows $T(t)$ (right
    scale) and the thick magenta line shows the energy recorded in the
    numerical simulation (left scale).
    The dashed black line shows the energy $E(t)$ calculated
    analytically (Appendix \ref{app:energyt}).
  }
  \label{fig:te1054}
\end{figure}

\medskip
Figure \ref{fig:te1054} shows a typical simulation result, for $\omega = 0.1$
and $\dtac = 10^{-4}$ chosen so that the magnitude of the temperature
modulation is much smaller that $T_0$, as in actual temperature modulated
calorimetry 
experiments. After a short transient, discussed below, the system reaches a
steady state in which the energy oscillates, with a phase shift with respect
to the temperature modulation.

The heat capacity of the system $C(\omega,t) = dE/dT$ can be derived
from the simulation results $C(\omega,t) =
[dE(t)/dt]\,/\,[dT(t)/dt]$. Separating the
fraction of $dE(t)/dt$ which is in phase with $dT(t)/dt$ one gets
$C'(\omega,t)$ while the fraction in quadrature with $dT(t)/dt$ gives
$C''(\omega,t)$, which correspond to the ``real'' and ``imaginary'' parts of 
$C(\omega,t)$ when one uses a complex notation.

\begin{figure}[h]
  \centering
  \includegraphics[width=7cm]{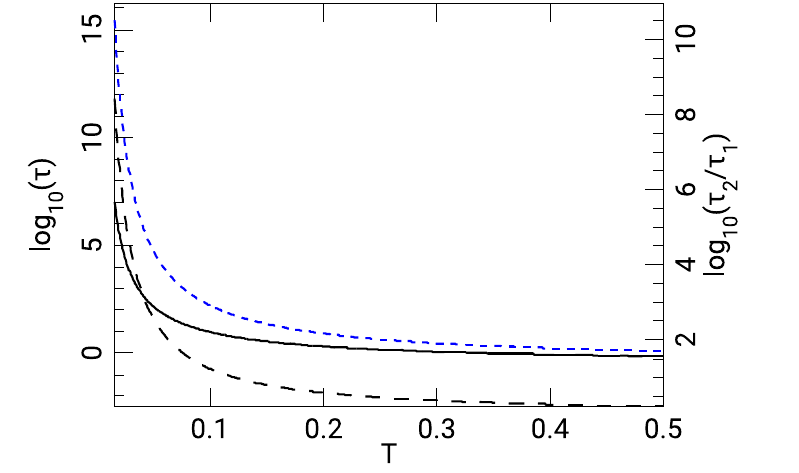}
  \caption{Relaxation times $\tau_1$ (black full line) and $\tau_2$
    (blue dotted line) versus temperature,
    and the ratio $\tau_2/\tau_1$ (black dashed line).
    The relaxation times are
    measured in the time unit defined by setting the prefactors $\omega_{ij} =
  1$ in Eq.~(\ref{eq:wij}).}
  \label{fig:tau}
\end{figure}

However, for the three-state
system, $C(\omega,t)$ can also be calculated analytically, as shown in
Appendices \ref{app:energyt} and \ref{app:heatcapacity}, which gives a much
better insight into the actual mechanisms which contribute to the dynamic
specific heat.
As the amplitude of the temperature modulation is small, the rates of the
thermally activated transitions can be expanded to first order around their
values at temperature $T_0$. It is convenient to introduce the deviations
$Q_i(t) = P_i(t) - P_i^{\mathrm{eq}}(T_0)$. In the general case these
deviations are not assumed to be small because the calculation also applies
for instance when we study a system which has been brought to $T_0$ after a
large temperature jump. The analytical solution amounts to solving a set
of two coupled equations for $dQ_1/dt$ and $dQ_2/dt$ which derive
from Eq.~(\ref{eq:dp1dt}) and from the corresponding
equation for $dP_2/dt$. The two equations can be written in a matrix form and
the solution is expressed on the basis of the two eigenstates of the
$2 \times 2$ matrix which relates the $Q_i$s and their time derivatives. Each
eigenmode has a relaxation time $\tau_i$ so that the calculation shows that the
dynamics of the model is governed by two relaxation times $\tau_1$ and
$\tau_2$, which depend on temperature.
Figure \ref{fig:tau} shows how the relaxation times depend on temperature for
the parameter set that we have chosen.
In the high temperature limit
  the rates of the transitions tend to unity according
  to Eq.~(\ref{eq:wij}), with the time unit defined by
setting  $\omega_{ij} =  1$, whereas, in the low
temperature range, 
$T \lesssim 0.03$, $\tau_1$ and $\tau_2$ grow by many orders of magnitude in a
narrow temperature range so that the three-state system exhibits a glass-like
behavior although it does not have a true glass transition.

At $T_0 = 0.125$ the relaxation times are
$\tau_1 = 5.217\;$t.u. and $\tau_2 = 49.11\;$t.u. corresponding to
eigenfrequencies $\omega_1 = 1.204\,$t.u.$^{-1}$ and $\omega_2 =
0.1279\,$t.u.$^{-1}$.
The dashed black line on Fig.~\ref{fig:te1054} shows the analytical result for
$E(t)$.
It exactly matches the magenta line showing the energy $E(t)$ deduced from
the numerical simulation, which indicates that the linear expansion of the
reaction rates is sufficient to accurately describe the dynamics of the
system. The agreement remains very good even if $\dtac$ is increased by one
order of magnitude to $\dtac = 10^{-3}$.

\begin{figure}[h]
  \centering
  \includegraphics[width=7cm]{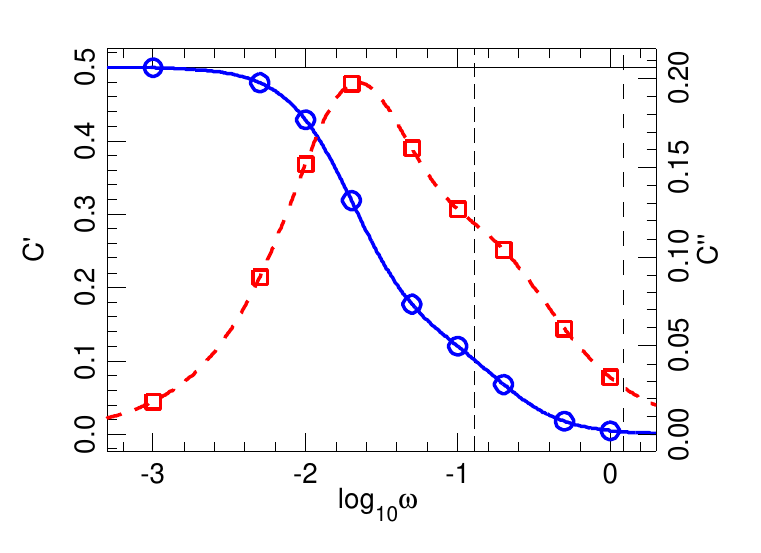}
  \caption{Heat capacities $C'(\omega)$ (blue full line) and $C''(\omega)$ (red
    dashed line) for the three-state system
    with the parameters that we have chosen,
    at temperature $T_0 = 0.125$. The lines result from the analytical
    calculations (Appendices \ref{app:energyt} and \ref{app:heatcapacity}). The
  points have been obtained from numerical simulations for different values of
$\omega$. The vertical dashed lines show the values of $\omega_1$ (long dash)
and $\omega_2$ (short dash). The black horizontal line
near the top of the graph shows the value of the
equilibrium heat capacity at temperature $T_0$ on the scale of $C'$.}
  \label{fig:comeg}
\end{figure}

\medskip
After the initial transient the energy reaches a steady oscillatory state and
the heat capacity depends on $\omega$ only. Temperature Modulated
calorimetry can get access to the spectrum of the characteristic times
of the energy transfers
within the degrees of freedom of sample by carrying a series of experiments
with different modulation frequencies. For the three-state model
that spectrum is known because the
analytical calculation determines the two relaxation times $\tau_1$, $\tau_2$
and how they influence the specific heat because it gives the functional form
of $C'(\omega)$ and $C''(\omega)$, plotted on 
Fig.~\ref{fig:comeg} for the model parameters that we have chosen. This figure
also shows individual points, deduced from a series of numerical simulations
with different values of $\omega$. They illustrate how one would build the
curves for $C'(\omega)$ and $C''(\omega)$ in a series of experiments.
The shape of the curves shows how the eigenfrequencies play the role
of cutoff frequencies for energy transfer. Above the highest frequency none of
the modes can be excited and $C(\omega)$ quickly drops to $0$. In the frequency
range $\omega_2 < \omega < \omega_1$ one of the two modes can still be excited
and therefore $C(\omega)$ still keeps a significant value. At very low
frequency, as expected $C'(\omega)$ tends to the equilibrium heat capacity
while $C''(\omega)$ tends to vanish.

\begin{figure}[h]
  \centering
  \includegraphics[width=7cm]{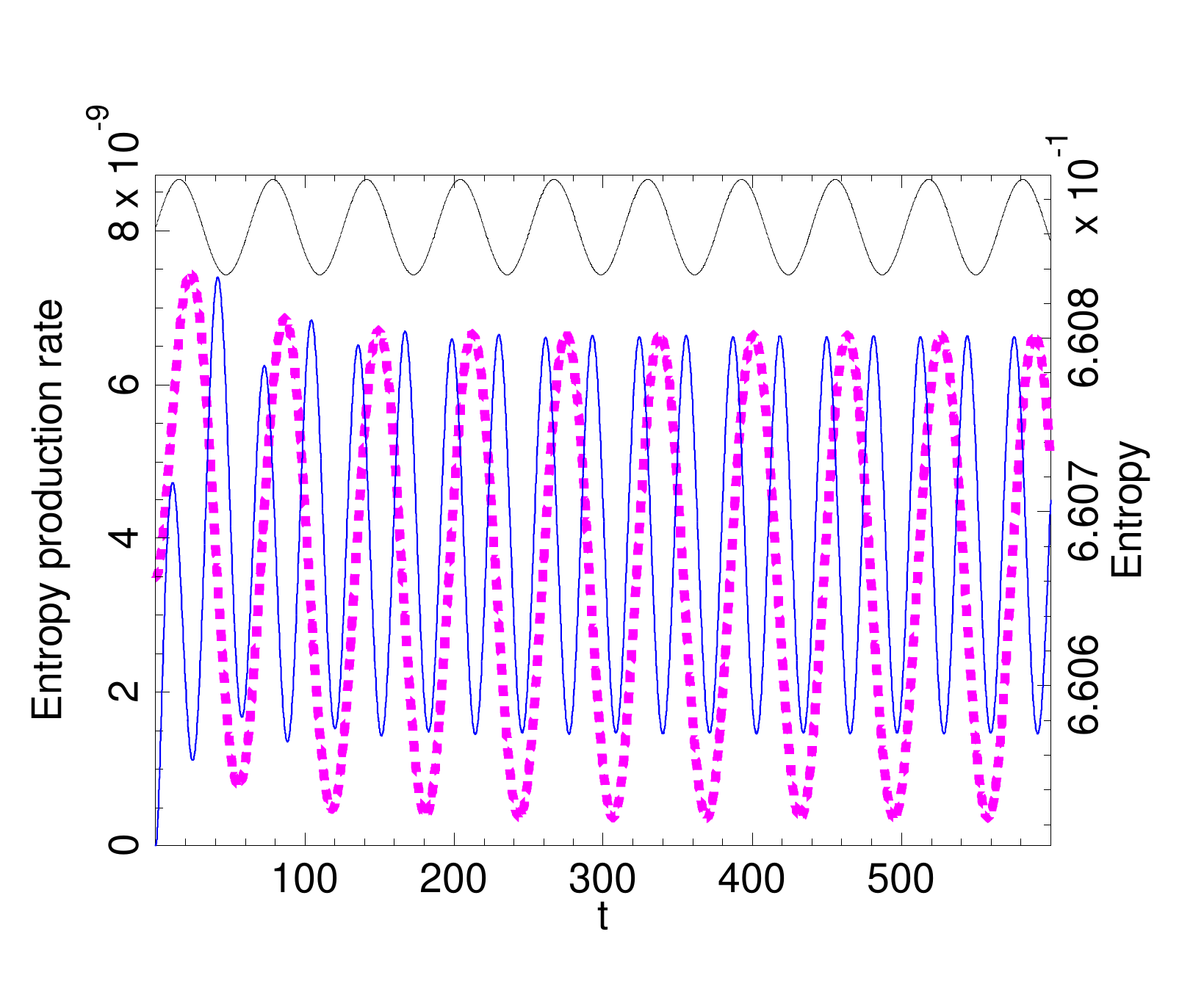}
  \caption{Entropy $S(t)$ (magenta dashed line, right scale)
    and entropy production rate
    $\mathcal{A}_S$ (blue full line, left scale)
    for the numerical experiment shown in Fig.~\ref{fig:te1054}.
    The sinusoidal function on top shows the variation
    of the temperature as a phase reference (no scale).}
  \label{fig:entropy}
\end{figure}

\smallskip
The physical meaning of the frequency-dependent heat capacity has been widely
discussed in the literature \cite{GARDEN-REVIEW}, and particularly its
so called ``imaginary part'', which appears as linked to some form of heat
dissipation. It actually describes the component of the response which is
in quadrature of phase with the temperature modulation due to delays caused
by energy transfers between the various energy states.
Using the three-state model the origin of this contribution to
the heat capacity can be related to the entropy production in an
out-of-equilibrium system. In a thermodynamic transformation, the variation
$\Delta S$ of the entropy includes two contributions. The contribution
$\Delta_{\mathrm{e}} S = \Delta E / T$
comes from the exchange of energy $\Delta E$ with
the environment. The second contribution $\Delta_{\mathrm{i}} S$ is associated to
internal transformations within the system. It is never negative
$\Delta_{\mathrm{i}} S \ge 0$ and vanishes for a reversible process.
For the three-state system, both simulations and analytical calculations can
determine the time evolution of the probabilities of occupation of the
metastable states $P_i(t)$, which were used in the calculation of
$C(\omega)$. Therefore we can calculate the entropy of the system
\begin{equation}
  \label{eq:entropy}
  S(t) = - \sum_{i=1}^3 P_i(t) \; \ln P_i(t) \; .
\end{equation}
The variation of entropy in an elementary transformation in which the
probabilities of occupation change by $dP_i$ is therefore
$dS = - \sum_{i=1}^3 \ln P_i \; dP_i$ (taking into account $\sum dP_i = 0$).
$d_{\mathrm{e}} S = (1/T)  \sum_{i=1}^3 E_i \; dP_i$ can be expressed in terms
of the equilibrium occupation probabilities at temperature $T$ as
$d_{\mathrm{e}} S = - \sum_{i=1}^3 \ln P^{\mathrm{eq}}_i \; dP_i$ so that the
entropy production rate is
\begin{equation}
  \label{eq:entropyprod}
  \mathcal{A}_S = \frac{d_{\mathrm{i}}S(t)}{dt} = \frac{dS(t)}{dt}
- \frac{d_{\mathrm{e}}S(t)}{dt} = - \sum_{i=1}^3 \ln \left(
\frac{P_i}{P^{\mathrm{eq}}_i} \right) \frac{dP_i}{dt} \; .
\end{equation}
It is shown in Fig.~\ref{fig:entropy} for the
transformation shown in Fig.~\ref{fig:te1054} during which a temperature
modulation is applied to the three-state system initially in
equilibrium at $T=0.125$. The entropy production rate is very
small (its maximum is of the order of $8 \, 10^{-9}$ while the entropy is of
the order of  
$6 \, 10^{-1}$) but nevertheless non-zero, and always positive as expected. It
shows two peaks per period of the temperature modulation, one when energy
rises and one when it decreases. It simply means that the modulation of the
temperature is too fast to allow the system to stay in equilibrium when
temperature and energy change (whatever the change, positive or negative) and
this leads to entropy production.
These out-of-equilibrium processes are
responsible for the $C''(\omega)$ term in the heat capacity, which has been
described as a ``loss term'' in some studies \cite{GARDEN-REVIEW}.
The frequency dependence of the maximum of the entropy production rate, which
rises to $2\,10^{-8}$ for $\omega = 1.0\,$t.u.$^{-1}$ and
drops to $1.2\,10^{-11}$ for $\omega = 10^{-3}\,$t.u.$^{-1}$
attests of the role of out-of-equilibrium phenomena in a
temperature modulated calorimetry experiment.

\medskip
Up to now we discussed the steady state of the simulated experiment
(Fig.~\ref{fig:te1054}). However the simulation shows a small decay of the
average energy before this steady state is reached, and this may seem
surprising because we started from an equilibrium state at temperature $T_0$
and imposed a small sinusoidal temperature modulation around $T_0$. Therefore
the average temperature has been maintained at $T=T_0$. Actually the energy
shift, due to this modulation is easy to understand by looking at the curve
showing how the equilibrium energy $E^{\mathrm{eq}}(T)$
varies versus the temperature of the system (Fig.~\ref{fig:ceequil}).
Around $T_0=0.125$, $E^{\mathrm{eq}}(T)$ varies non-linearly versus
temperature and is curved downwards. As a result, for a system oscillating
along this curve between $T_0 - \dtac$ and $T_0 + \dtac$ the average energy is
expected to be {\em smaller than} $E^{\mathrm{eq}}(T_0)$. This simple
qualitative explanation can be checked by performing a similar simulation at a
temperature $T'_0$ around which $E^{\mathrm{eq}}(T)$ is curved upwards. 
This is the case for $T'_0 = 0.05$ as shown in Fig.~\ref{fig:ceequil} and
indeed a simulation shows that the average energy moves up when a sinusoidal
modulation of the temperature around $T'_0$ is imposed. Of course, as
discussed above, the system subjected to such a sinusoidal modulation is not
in equilibrium. Thus the equilibrium $E^{\mathrm{eq}}(T)$ cannot provide an
accurate evaluation of the average energy shift in the presence of the
modulation $\tac$. To get such an evaluation one has to calculate the actual
evolution of the energy versus time, as in Appendix \ref{app:energyt}, which
is plotted as a dashed black line in Fig.~\ref{fig:te1054}.
The figure shows that
this analytical calculation matches the simulation results (thick
magenta line on
Fig.~\ref{fig:te1054}). The role of a sinusoidal temperature modulation to 
modify the average value of some property of a material
has already been
observed experimentally \cite{JOHARI1999} for the dielectric relaxation and
thermodynamic properties of polymers. The understanding of these phenomena,
further refined in \cite{WANG-JOHARI}
involves exactly the same mechanism that we
exhibited for the three-state system: the nonlinear change of this property when
temperature varies. In some cases experiments show that the effect can become
large, and this is also the case for the three-state model around $T_0 = 0.04$
because, below this temperature $E^{\mathrm{eq}}(T)$ becomes almost flat,
while above $T_0 = 0.04$ it starts to raise significantly. In this temperature
range the calculation of Appendix \ref{app:energyt} loses accuracy because
expanding the rates of the thermally activated transitions to first order in
$\tac$ is not enough. Higher order terms, introducing further nonlinearities,
start to play a significant role.

\subsection{Time-dependent heat capacity during the aging of a sample at
  constant temperature}
\label{sec:aging}

Understanding aging in glasses is a challenge for theoretical
physics. Experiments are generally made by following the properties of a glass
while it is cooled fast enough to prevent it from reaching
equilibrium. In this case aging results from the combined effect of the
temperature change and intrinsic phenomena within the glass. This
makes the analysis more complex.
Using a modulated temperature as in
Eq.~(\ref{eq:Tmodul}) allows measurements of thermodynamic
properties while the glass ages at constant temperature because
$\tac \ll T_0$ can be chosen so small that it has a negligible influence on
the aging. Moreover $\omega$ can be varied to follow specific time scales
during aging. The three-state system, which has a spectrum of
fluctuations which is richer than a simple relaxation, is a interesting case
to study how aging can lead to a time dependent heat capacity of a glass
during aging.

\smallskip
\begin{figure}[h]
  \centering
  \begin{tabular}{c}
    \textbf{(a)} \\
    \includegraphics[width=7cm]{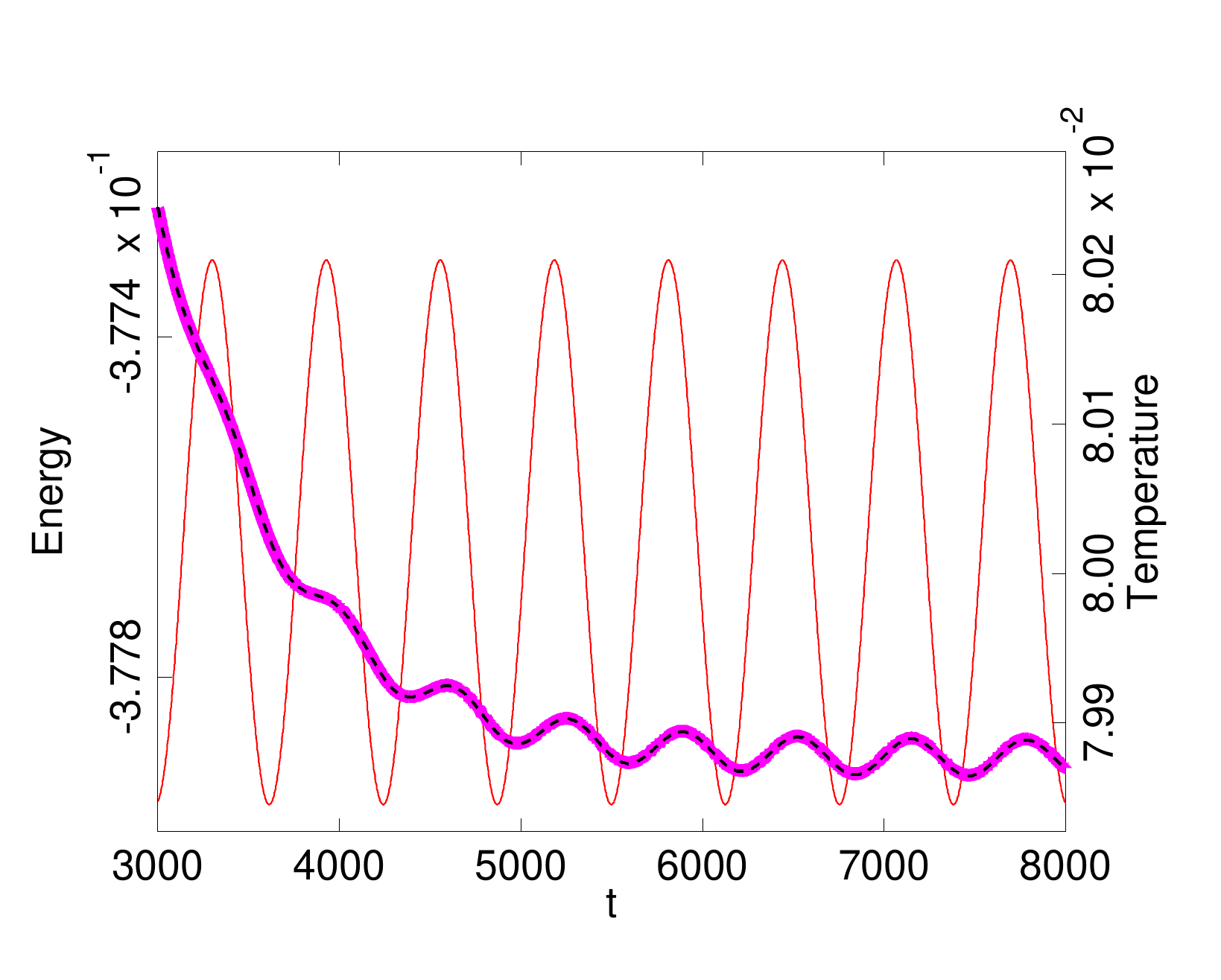} \\
     \textbf{(b)}\vspace{-0.4cm} \\
    \includegraphics[width=7cm]{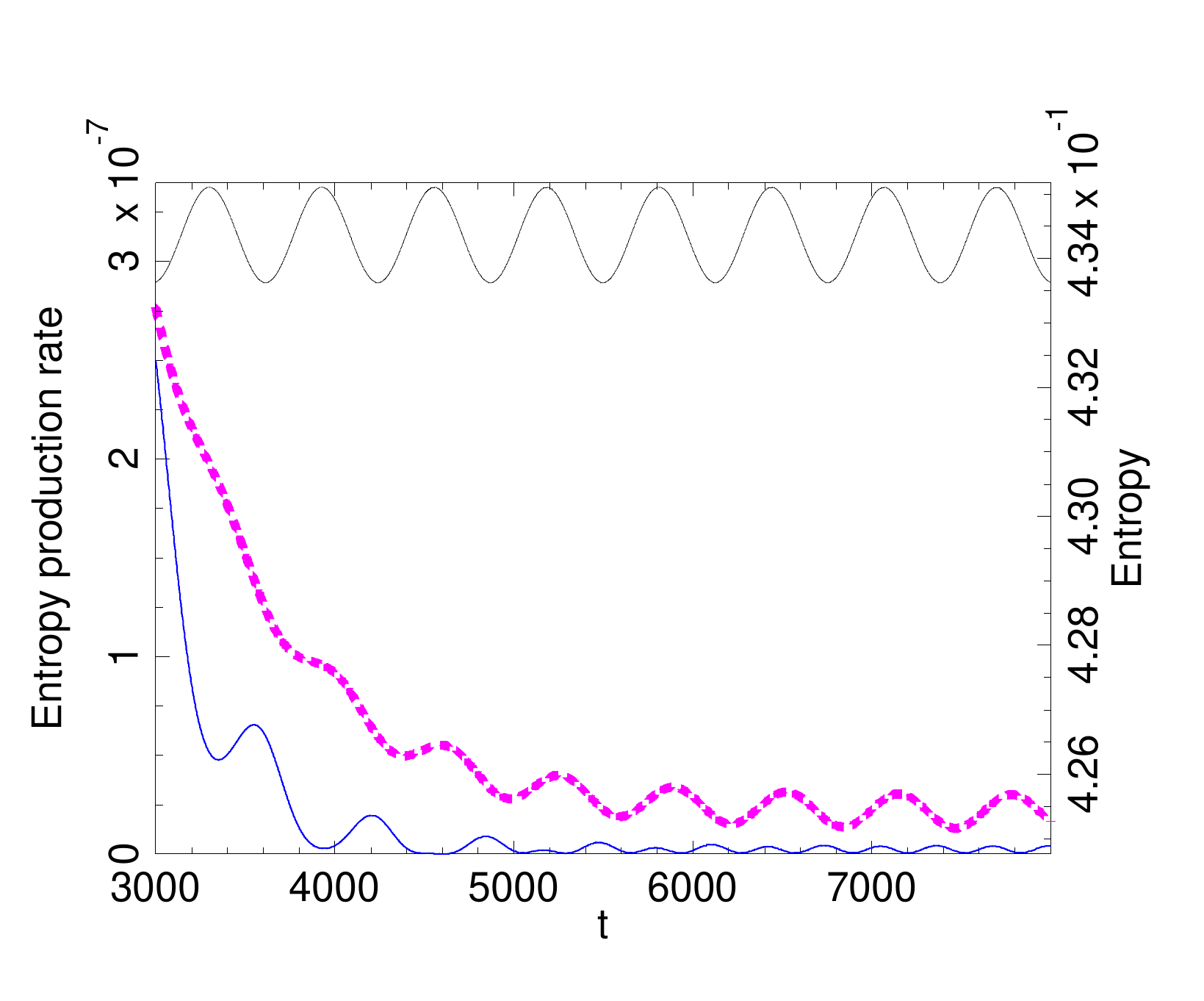}
  \end{tabular}
  \caption{Time evolution of the thermodynamic properties of the three-state
    system during aging at $T_0 = 0.08$, with a modulation
    $\tac = \dtac \sin \omega t$ ($\omega = 10^{-2}\,$t.u.$^{-1}$,
    $\dtac = 2\,10^{-4}$) after a temperature jump from
    equilibrium at $T=0.3$. (a) Energy (thick magenta full line
    from simulation results, dotted black line
    from analytical calculation - left scale) and temperature
    (thin red full line, right
  scale). (b) Entropy (magenta dashed line, right scale)
  and entropy production rate (blue full line,
  left scale). The sinusoidal curve shows temperature as
  a reference (scale not indicated). }
  \label{fig:aging1071}
\end{figure}
\medskip

\begin{figure}[h]
  \centering
  \begin{tabular}{c}
    \textbf{(a)} \\
    \includegraphics[width=7cm]{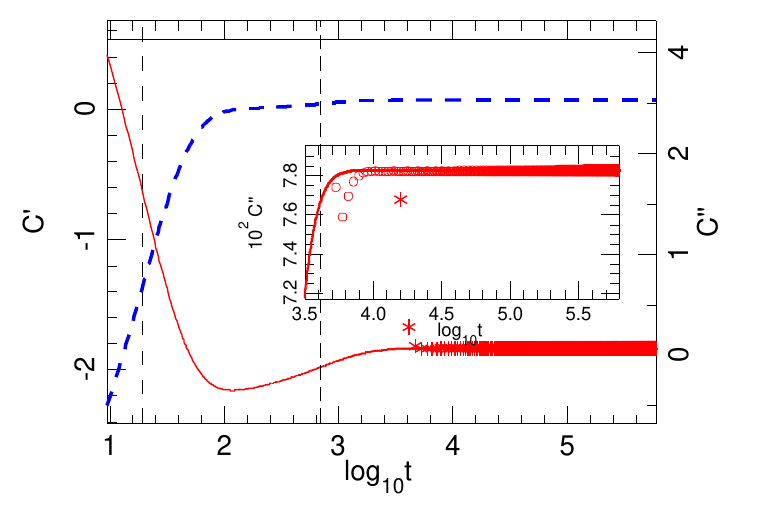} \\
     \textbf{(b)} \\
    \includegraphics[width=7cm]{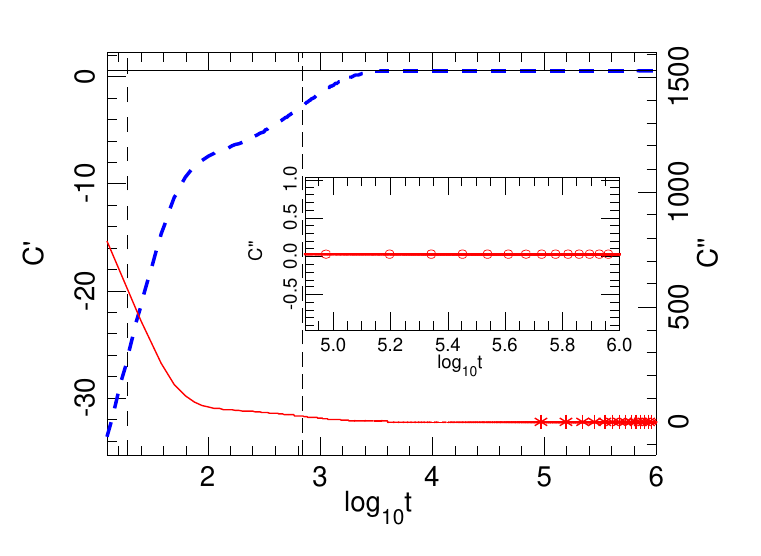}
  \end{tabular}
  \caption{Dynamic heat capacity $C'(\omega)$ (blue dashed line,
    left scale) and
    $C''(\omega)$ (red full line, right scale) of the three-state
    system, versus time after a temperature jump from $T=0.3$ to
  $T_0 = 0.08$. The horizontal full line shows the value of the equilibrium
  heat capacity $C^{\mathrm{eq}}(T_0)$ on the scale of $C'(\omega)$. The dashed
  vertical lines show the values of $\tau_1$ (long dash)  and $\tau_2$ (short
  dash). The stars (circles
  in insets)
  show the value of $C''$ deduced from the entropy production
  according to \cite{GARDEN-RICHARD} for $t > 2 \pi / \omega$ (
  condition imposed by the
  integration over a cycle of the modulation to compute
  $C''$). (a) results for $\omega = 10^{-2}$t.u.$^{-1}$,
(b) results for $\omega = 10^{-4}$t.u.$^{-1}$.}
  \label{fig:comegaging}
\end{figure}

Figure \ref{fig:aging1071} shows the behavior of energy, entropy, and entropy
production during the early stage of aging of the three-state system after a
temperature jump. In this simulation the model studied above (with the same
parameters) was kept in equilibrium at $T = 0.3$ for a short time and then its
temperature was abruptly changed to $T_0 = 0.08$ while we followed its
properties by adding a small modulation with frequency $\omega =
10^{-2}\,$t.u.$^{-1}$ and
amplitude $\dtac = 2\,10^{-4}$. At $T_0 = 0.08$ the relaxation times are
$\tau_1 = 19.09\;$t.u. ($\omega_1 = 0.329\,$t.u.$^{-1}$) and $\tau_2 =
690.9\;$t.u. ($\omega_2 = 0.91\,10^{-2}\,$t.u.$^{-1}$).
The temperature jump is followed by
a strong relaxation of the energy, with a large entropy production. Figure
\ref{fig:aging1071} starts when this initial stage is almost over because the
variations in energy and entropy at very short time are so large that they
hide the effect of the temperature modulation (without preventing nevertheless
the determination of $C(\omega)$ as shown in Fig.~\ref{fig:comegaging}).
The last stage of the relaxation is visible on
Fig.~\ref{fig:aging1071}. The variation versus time of the in-phase
$C'(\omega)$ and out-of-phase $C''(\omega)$ components of the dynamic heat
capacity are shown in Fig.~\ref{fig:comegaging} using a logarithmic scale for
time to display the properties at various time scales more clearly.

\medskip
The first important point to notice is that the dynamic heat capacity,
which corresponds to the response of the system at a particular
frequency (determined for instance by Fourier transform in the analysis of
experimental data) actually shows the strong relaxation phenomena that
follow the heat jump, even if their time scale does not match the period of
the modulation. This looks surprising at first examination, but this should
actually be expected because the relaxations correspond to evolutions within
the sample, which modify its thermal response.
The physical mechanism of this phenomenon, which may
  lead to a large time-dependence of the dynamic heat capacity in a strongly
  out-of-equilibrium system, is the
following. Because the rates of the thermally activated transitions
$ W_{i \to j}$ given by Eq.~(\ref{eq:wij}) depend on temperature, they are
modulated by the signal $T_{\mathrm{ac}}(t)$. In a first order expansion
they include a contribution proportional to $T_{\mathrm{ac}}(t)$. The
master equations giving $dP_i/dt$ contain products of $P_j$
by $W_{i \to j}$, and therefore $dP_i/dt$ contains terms proportional
to $P_j \times T_{\mathrm{ac}}(t)$. They show up in the term $C_4$ of
Eq.~(\ref{eq:defci}) of Appendix \ref{app:energyt}. This coupling of the
temperature oscillations with the departure of the probabilities
from equilibrium, which follows a temperature jump or a very fast cooling,
gives rise to a strong time-dependence of the dynamic heat capacity.
Therefore it should be expected that $C'(\omega)$, $C''(\omega)$, which
measure the response of the system to the temperature modulation
$T_{\mathrm{ac}}(t)$, follow the relaxation of the system after a large
temperature jump, or a very fast cooling. The calculation presented in
Appendix \ref{app:energyt} gives a quantitative evaluation of this
effect which can cause a strong time-dependence of the dynamic heat capacity.
Linear response theory \cite{NIELSEN},
which treats systems near equilibrium, neglect this
coupling  and therefore finds that the dynamic specific heat depends
on the frequency only, and not on time.

Such large variations of the dynamic heat capacity versus time appear in
Fig.~\ref{fig:comegaging}, which shows that, in the
early stage of the evolution, $C'(\omega)$ can even be {\em negative}.
Negative heat capacities have been measured in temperature scans for glasses
strongly out of equilibrium \cite{THOMAS1931}. Temperature Modulated
calorimetry can detect 
a similar phenomenon during aging at constant temperature after a large
temperature jump.
Moreover, for dynamic heat capacity, negative values are not
surprising, even close to equilibrium, because the theoretical analysis shows
that the dynamic heat capacity does not share with the equilibrium heat
capacity the property of being always positive \cite{FIORE}.

\begin{figure}[h]
  \centering
  \begin{tabular}{c}
  \textbf{(a)} \\
  \includegraphics[width=7cm]{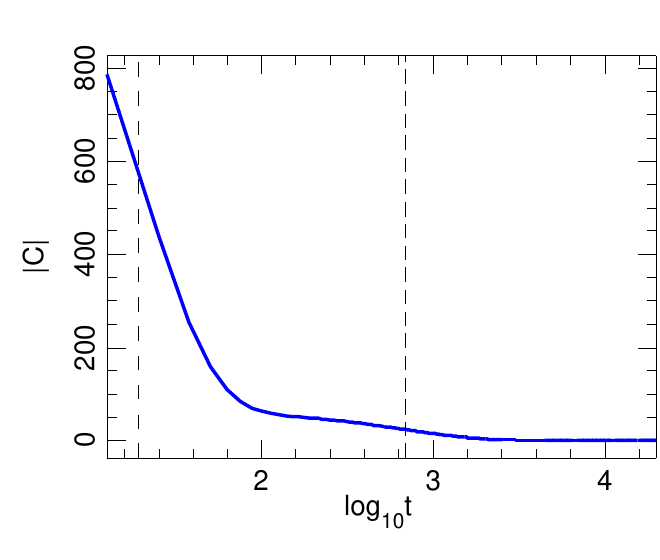} \\
    \textbf{(b)} \\
  \includegraphics[width=7.0cm]{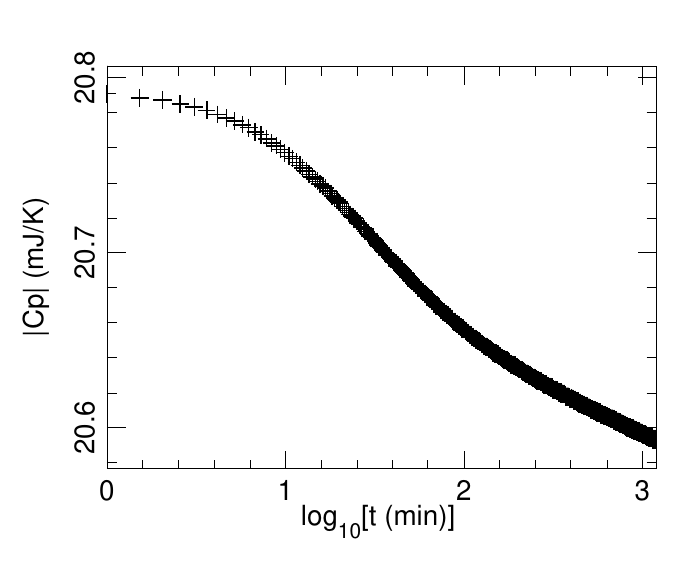}
    \end{tabular}
  \caption{a) Modulus of the dynamic
    heat capacity versus time (logarithmic
    scale) for the three-state system aging after a temperature jump
    from $T=0.3$ to $T_0 = 0.08$ determined with a temperature modulation at
  $\omega = 10^{-4}$t.u.$^{-1}$ (corresponding to the results shown in
  Fig.~\ref{fig:comegaging}-b). The dashed vertical lines show the values of
  $\tau_1$ and $\tau_2$ at temperature $T_0$.
  b) Experimental result for the aging of a PVAC sample
  at $24^{\circ}$C after cooling from $90^{\circ}$C at $-1.2^{\circ}$C/min.
  The modulus of the heat capacity $C_p$ was measured at constant temperature
  using a temperature modulation at $0.1954\,$Hz.
  (Data from \cite{LAARRAJ}).
}
  \label{fig:ClogtPVAC}
\end{figure}
As shown in Fig.~\ref{fig:ClogtPVAC}-a, when the three-state system ages after
a sharp cooling, the modulus of its dynamic heat capacity at fixed frequency
may show a large variation with time. This phenomenon takes place although
nothing has been modified in the model itself or its parameters. It is due to
the large relaxation phenomena following a heat jump discussed above. Actual
measurements on aging glasses have also detected strong time dependencies of
the modulus of the dynamic heat capacity $|C_p|$ at fixed frequency during
aging. Figure~\ref{fig:ClogtPVAC}-b shows experimental data for PVAC after a
fast cooling \cite{LAARRAJ}. In a real system, like PVAC, the phenomena are
more complex than for the theoretical three-state system, and this simple
model would not be sufficient to fit the actual data, but it suggests
nevertheless a pathway to understand the experiments.
The results for the
three-state system clearly show two stages in the aging, associated to the two
relaxation times $\tau_1$ and $\tau_2$. PVAC has a richer spectrum of
relaxation times and one cannot readily identify them from
the experimental curve of Fig.~\ref{fig:ClogtPVAC}-b. Nevertheless
the results suggest at least a very fast relaxation of the order of a few
minutes, a second range of relaxation times around 30 minutes, and the
presence of longer relaxation times, extending beyond hundreds of minutes
as the relaxation is not over at the end of the 1000-minute experiment.
The decay with time of the dynamic heat capacity has also been observed by
temperature modulated calorimetry for another glassy sample,
triphenylolmethane triglycidyl ether
\cite{TOMBARI2002}. In the discussion of the results, this study considered
various possible explanations, such as an evolution of the relaxation times in
the sample or indirect effects of structural relaxation leading to a density
increase. The simple three-state model cannot attempt to detect subtle effects
which are sample-specific, but it shows however that relaxations, with
constant characteristic times, are sufficient to generate such a decay of the
heat capacity versus time, possibly allowing a simpler analysis of the
observations of Tombari et al. \cite{TOMBARI2002}.

\bigskip
A theory of the frequency dependent specific heat has been proposed in
\cite{NIELSEN}. It is based on the fluctuation-dissipation theorem and
only applies in the vicinity of equilibrium. Our approach, which solves the
equations for the time evolution of the occupation probabilities
shows that the response  of a system very far from equilibrium to a modulation
of its temperature can be more complex and includes a contribution coming from
the relaxation of the system towards equilibrium. When the system approaches
equilibrium, our results converge to the expression given in
\cite{NIELSEN}, but they are more general. The drawback of our approach
is that the dynamic equations are solved in the
context of a particular model.

\medskip
Another general analysis of temperature modulated calorimetry has been
proposed in 
\cite{GARDEN-RICHARD}. It focuses on the origin of the ``loss term''
$C''(\omega)$ and shows that, in the vicinity of equilibrium, it can be
quantitatively related to the entropy production integrated over one cycle of
the modulation, $\overline{{\cal A}_s}$, by
\begin{equation}
  \label{eq:centropy}
  \overline{{\cal A}_s} = \pi \left(\frac{\dtac}{T_0}\right)^2 C'' \; .
\end{equation}
The value of $C''$ deduced from Eq.~(\ref{eq:centropy}) is plotted on
Fig.~\ref{fig:comegaging} (see insets for a magnified scale).
It converges towards $C''(\omega)$ for all
$\omega$ when aging has been long enough to allow the system to come
sufficiently close to equilibrium to ensure the validity of the
Eq.~(\ref{eq:centropy}) \cite{GARDEN-RICHARD}. The results shown on
Fig.~\ref{fig:comegaging}-b is however trivial because the modulation is so slow
($\omega = 10^{-4}\,$t.u.$^{-1}$) that it probes quasi-equilibrium properties
of the system, when entropy production has vanished as well as $C''$.
Figure~\ref{fig:comegaging}-a is more interesting because it shows the
quantitative validity of the approach of \cite{GARDEN-RICHARD}.

\medskip
Figure \ref{fig:comegaging}, which shows the evolution of the dynamic heat
capacity for two values of $\omega$, suggests that such measurements could
allow a study of the various time scales involved when a system ages.
The two characteristic times which control the aging for the
three-state system are marked on this figure.
As discussed above, the large variations of $C'(\omega)$
and $C''(\omega)$ provide an indirect view of the relaxations toward
equilibrium. The shape of $C'(\omega)$ gives a first insight on the
relaxations in the system because it exhibits two quasi-linear segments
centered around $t = \tau_1$ and $t=\tau_2$. In a system
with a more complex spectrum of aging timescales,
performing experiments for different values of $\omega$ may give a
quantitative information on this spectrum. The simulation
with $\omega = 10^{-4}\,$t.u.$^{-1}$
probes the evolution of the system up to timescales
larger than $\tau_2$ and therefore a quasi-equilibrium situation. This is why,
in the long term $C'(\omega)$ tends to $C^{\mathrm{eq}}(T_0)$. Conversely
for $\omega = 10^{-2}\,$t.u.$^{-1}$,
even in the large time limit $C'(\omega)$ stays well
below $C^{\mathrm{eq}}(T_0)$ because the time scale of the modulation does not
allow the system to relax the degrees of freedom which evolve with
characteristic time $\tau_2$. Thus studying
\begin{equation}
  \label{eq:diffcomeg}
  \Delta C(\omega) = C^{\mathrm{eq}}(T_0) - \lim_{t \to \infty} C'(\omega,t)
\end{equation}
versus $\omega$ provides some measure of the relaxation times that govern
aging in the system and of the magnitude of the contribution to the dynamic
heat capacity of the degrees of freedom associated to each of these relaxation
times. 

\subsection{Investigating the Kovacs effect by temperature modulated calorimetry}
\label{sec:kovacs}

The Kovacs effect is an interesting effect which points out the peculiarities
of out-of-equilibrium systems \cite{KOVACS}.
To our knowledge it has not been investigated
by temperature modulated calorimetry although such experiments could provide
useful insights 
on its mechanism. It was observed in a series of two experiments. First a
glass sample is slowly cooled to record the variation of its volume versus
temperature $v^{\mathrm{eq}}(T)$ in quasi-equilibrium. In a second experiment,
the sample is abruptly cooled from an initial temperature $T_1$ to a
low temperature $T_2$ in the vicinity of the glass transition
temperature $T_g$. The sample is let to age at $T_2$. Its volume slowly
decreases. When the volume has reached the value $v_0$ that it would have at
equilibrium at some temperature $T_0 > T_2$, $v_0 = v^{\mathrm{eq}}(T_0)$, the
temperature of the sample is abruptly switched from $T_2$ to $T_0$. At this
point the volume and temperature of the sample are the same as if it
was in equilibrium at temperature $T_0$. However this state has not be
reached by a quasi-equilibrium trajectory. Therefore the sample is {\em not} in
equilibrium. Kovacs observed that, when it is maintained
at $T_0$, its volume
starts to increase, and then decays until the system slowly reaches
equilibrium at $T_0$ with volume $v^{\mathrm{eq}}(T_0)$.

\begin{figure}[h]
  \centering
  \includegraphics[width=7cm]{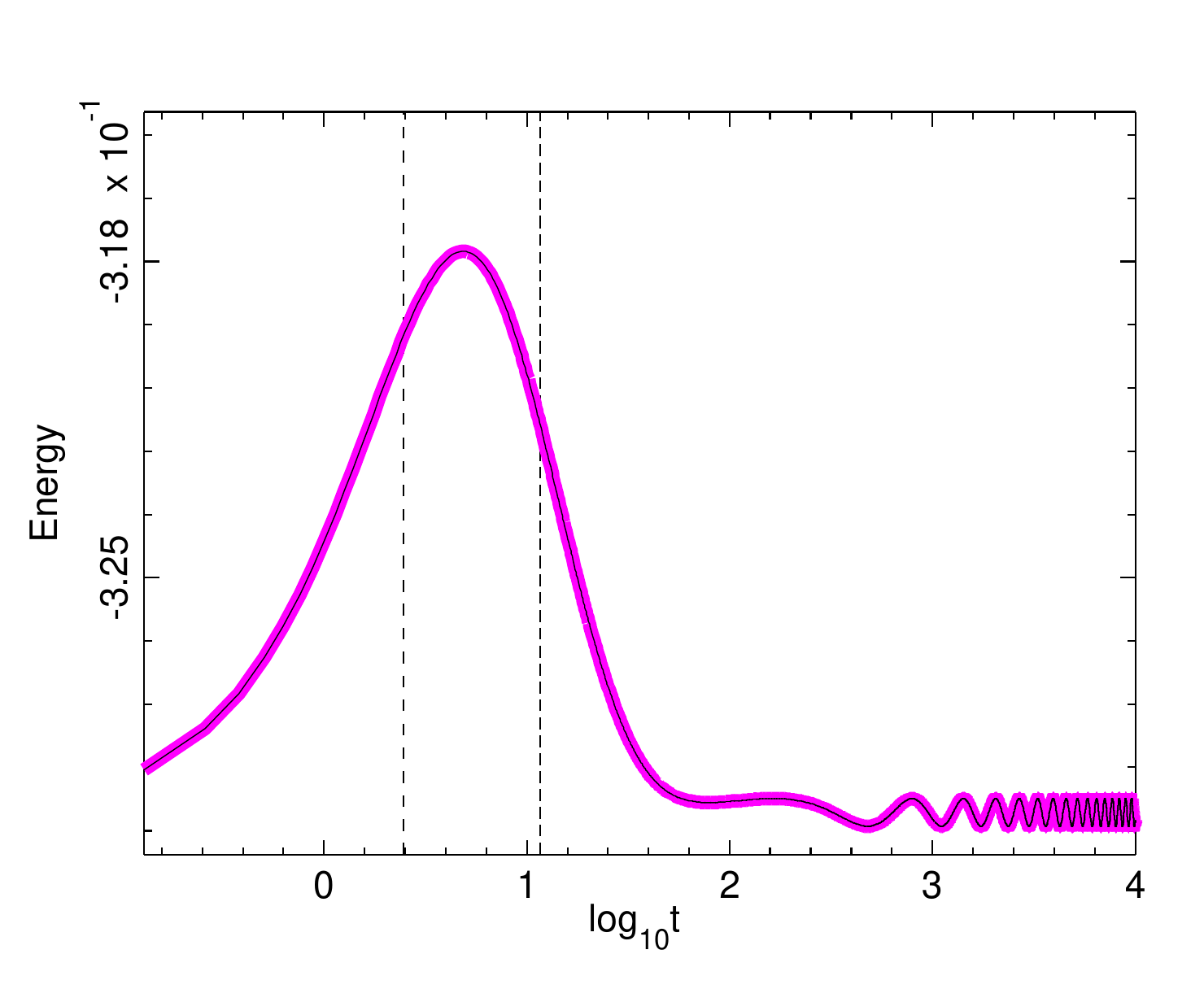}
  \caption{Energy versus time for a Kovacs scan with the three-state
    system. After a sharp temperature jump from $T_1=0.3$ to $T_2=0.02$ the
    model was let to age at $T_2$. Its energy decreases and, when it had
    reached the value $E^{\mathrm{eq}}(T=0.17856)$, the temperature was
    abruptly switched to $T = T_0 + \tac$ with $T_0=0.17856$ and
  $\tac = \dtac \sin \omega t$, $\dtac = 8\,10^{-4}$, $\omega =
  10^{-2}\;$t.u.$^{-1}$. The figure shows the energy of the system versus the
  time (in logarithmic scale) from the temperature jump from $T_2$ to $T_0$.
The thick magenta line shows the energy recorded during the numerical
simulation, while the thin black line, along the magenta line, shows the
result of the analytical calculation of the energy following the process
presented in Appendix \ref{app:energyt}, starting from the initial state
reached at the end of the aging period at $T_2$. The black vertical dashed
lines mark the relaxation times of the three state system at temperature
$T_0$.} 
  \label{fig:ekovacs}
\end{figure}

\medskip
For the three-state system, a volume is not defined, but nevertheless 
the Kovacs effect can be observed by following the energy versus time \cite{PG}.
Figure \ref{fig:ekovacs} shows an example from a numerical
simulation. From an equilibrium state at $T_1 = 0.3$, the three-state-system
(using the same model parameters as above in Sec.~\ref{se:spectrumequil}) was
abruptly switched to $T_2 = 0.02$ and let to age for $1.8\,10^7$t.u.
Its energy decreases slowly during this aging at very low temperature, and, at
the end of the aging period it has reached $E^{\mathrm{eq}}(T=0.17856)$.
Then the system is abruptly switched from $T_2 = 0.02$ to $T_0 = 0.17856$ and
we monitor its energy versus time in the presence of a small temperature
modulation $\tac$ with $\dtac = 8\,10^{-4}$ and various values of $\omega$.
Figure \ref{fig:ekovacs} shows an example for $\omega = 10^{-2}\;$t.u.$^{-1}$.
It exhibits the same ``Kovacs hump'' as the one observed by Kovacs for the
volume of a glass \cite{KOVACS}. Although, at the beginning of the scan at
temperature $T_0$ the system has the energy $E^{\mathrm{eq}}(T=0.17856)$,
while it ages at $T_0$ its energy rises significantly before coming back to
its equilibrium value. Figure \ref{fig:ekovacs} shows that the maximum occurs
at a time which is intermediate between the two relaxation times
$\tau_1$ and $\tau_2$ of the three-state model at temperature $T_0$. This
suggests that the shape of the Kovacs hump is governed by the relaxation
spectrum of the sample. It would be interesting to check this experimentally
with a glassy sample. Temperature Modulated calorimetry, which probes this
spectrum should 
be the appropriate technique. To test this idea, we have investigated the
evolution versus time of $C'(\omega)$ and $C''(\omega)$ during a Kovacs scan
of the three-state system.

\begin{figure}[h]
  \centering
  \begin{tabular}{c}
    \textbf{(a)} \\
    \includegraphics[width=7cm]{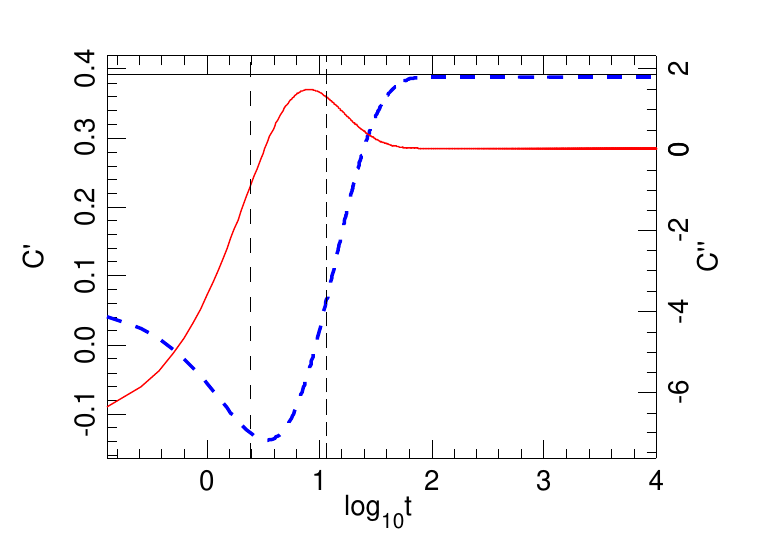} \\
     \textbf{(b)} \\
    \includegraphics[width=7cm]{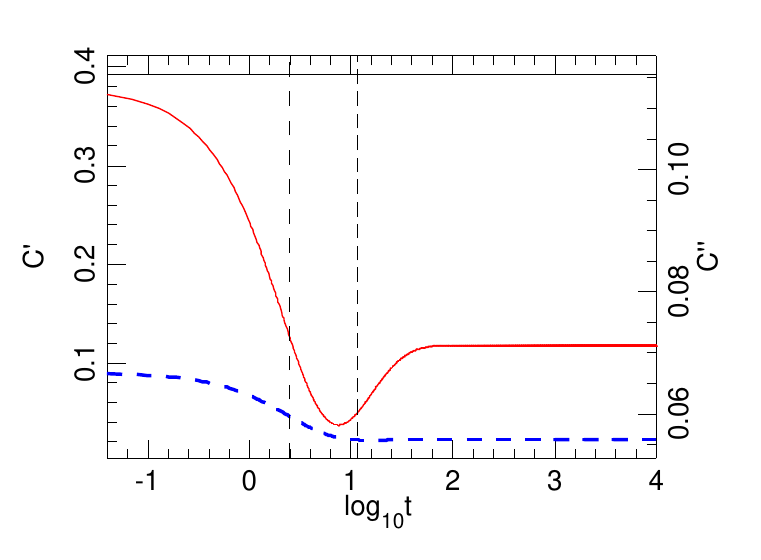}
  \end{tabular}
  \caption{Dynamic heat capacity $C'(\omega)$ (blue dashed line,
    left scale) and
    $C''(\omega)$ (red full line, right scale) of the three-state
    system, versus time during the Kovacs scan shown in
    Fig.~\ref{fig:ekovacs}. 
    The horizontal black line shows the value of the equilibrium
  heat capacity $C^{\mathrm{eq}}(T_0)$ on the scale of $C'(\omega)$. The dashed
  vertical lines show the values of $\tau_1$ and $\tau_2$. 
  (a) results for $\omega = 10^{-2}\,$t.u.$^{-1}$,
  (b) results for $\omega = 1\,$t.u.$^{-1}$.}
  \label{fig:ckovacs}
\end{figure}

Figure \ref{fig:ckovacs} shows the result for two values of $\omega$.
As shown in Fig.~\ref{fig:ekovacs}, the analytical calculation of the energy
versus time, as discussed in Appendix \ref{app:energyt}, matches the variation
recorded in the numerical simulation. This allows us to use the heat
capacity calculated analytically, which is more accurate than relying on a
numerical treatment of the simulation results, especially for a case where the
heat capacity depends on time.  
Figure \ref{fig:ckovacs} shows that both
$C'$ and $C''$ vary significantly with time. This is not surprising because we
have shown in Sec.~\ref{sec:aging} that, although temperature modulated
calorimetry records 
the value of the heat capacity at a specific frequency, the values of
$C'(\omega)$ and $C''(\omega)$ also reflect the relaxations which occur within
the sample. However the recorded signal is filtered in frequency because it is
detected through the response to an oscillatory driving. {\em This is what
makes temperature modulated calorimetry particularly attractive to study the
Kovacs 
effect.} On one hand it does reflect the evolutions which occur within the
sample while it ages during the Kovacs scan, as attested by the large
variations of $C'$ and $C''$ which take place in the same time scale as the
variation of the energy of Fig.~\ref{fig:ekovacs}. But, on the other hand, 
the variation of $C'(\omega)$ and $C''(\omega)$ give a quantitative
picture of the time scales at which the sample evolves.
The relaxation frequencies for the model at temperature $T_0$ are
$\omega_1 = 2 \pi / \tau_1 = 2.5427\;$t.u.$^{-1}$ and
$\omega_2 = 2 \pi / \tau_2 = 0.54193\;$t.u.$^{-1}$. Figure \ref{fig:ckovacs}-a
shows the dynamic heat capacity measured with a frequency $\omega =
10^{-2}\;$t.u.$^{-1}$, which probes the dynamic of the system at times
scales longer than its internal time scales. This is clear from
the results because
the value of $C'(\omega)$ which starts with a value well below the equilibrium
specific heat of the sample at $T_0$, $C^{\mathrm{eq}}(T_0)$, and even
temporary drops below $0$, finally reaches  $C^{\mathrm{eq}}(T_0)$ in the long
term, which indicates that the system had time to reach
a quasi-equilibrium on the time scale which is probed.
Figure~\ref{fig:ckovacs}-b shows the heat capacity
measured with a frequency $\omega' = 1\;$t.u.$^{-1}$,
$\omega_2 < \omega' <\omega_1$, 
which probes timescales larger than $\tau_1$ but only those smaller
than $\tau_2$. As a
result $C'(\omega')$ exhibits a
significant time-dependence, which indicates that the evolution within the
sample contains some
degrees of freedom which evolve faster than $2 \pi / \omega'$. However, in the
long term $C'(\omega')$ does not reach  $C^{\mathrm{eq}}(T_0)$. This indicates
that there are other degrees of freedom which are slower than $2 \pi /
\omega'$. Another simulation with $\omega'' = 4\;$t.u.$^{-1}$, i.e.
$\omega'' > \omega_1 > \omega_2$ finds that both $C'(\omega'')$
and $C''(\omega'')$
stay very small during the whole simulation, indicating that all degrees of
freedom are slower than $2 \pi / \omega''$. These results show how a study
of the dynamic heat capacity versus $\omega$ for an
experimental Kovacs scan might
clarify the role of the thermal relaxation spectrum in the shape of the Kovacs
hump of a glass.

\section{Scanning Calorimetry versus Temperature Modulated
  Scanning Calorimetry}
\label{sec:tmdsc}

\subsection{Experimental results}
\label{sec:experimental}

\begin{figure}[h]
  \centering
  \includegraphics[width=8.5cm]{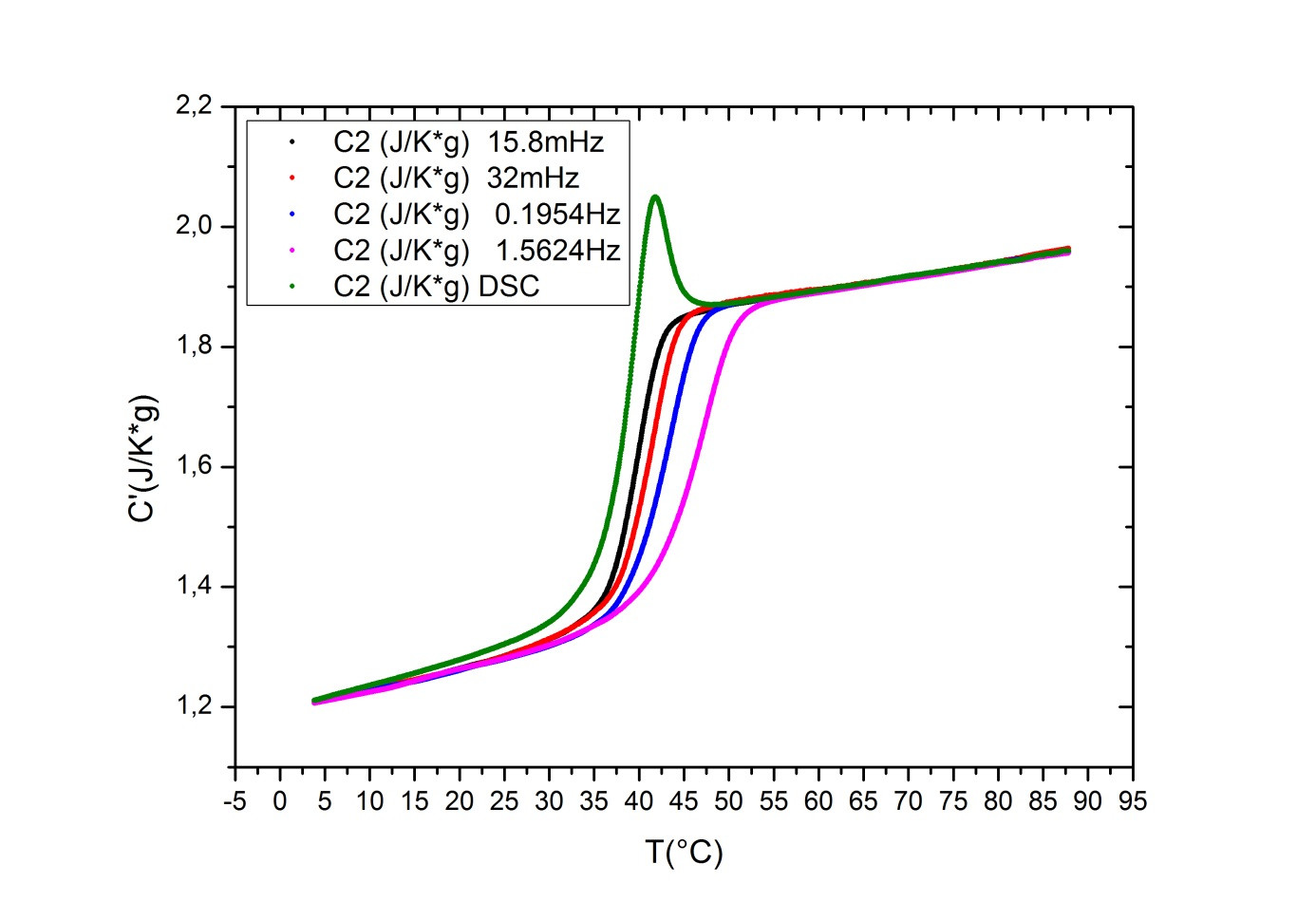}
  \caption{Comparison between scanning
    calorimetry results and temperature modulated calorimetry experiments for
    a PVAC 
    sample after fast cooling (Data from \cite{LAARRAJ}). 
    The sample was cooled from $90^{\circ}$C down to $4^{\circ}$C, with a
    cooling rate of $-1.2^{\circ}$C/minute.
    The green curve, with a maximum around $T=40^{\circ}$C, shows the result
    of a standard Differential Scanning Calorimetry (DSC)
    experiment during which
    the sample is heated at the rate of $+1.2^{\circ}$C/minute. The other
    curves show $C'(\omega)$ determined from temperature modulated scanning
    calorimetry (MTSC), with a temperature $T(t) = T_0(t) + \tac(t)$. $T_0(t)$
    is a temperature ramp with heating rate $+1.2^{\circ}$C/minute, and
    the frequency of the modulated part, $\nu = \omega / 2 \pi$, has the values
    $15.8\,10^{-3}$Hz (for the curve with the steepest rise), $32\,10^{-3}$Hz,
    $0.1954\,$Hz and $1.5624\,$Hz for the curve with the slowest rise.
  }
  \label{fig:pvac}
\end{figure}

Figure \ref{fig:pvac} points out a qualitative difference between
Differential Scanning Calorimetry (DSC) measurements and $C'(\omega)$
determined by a Modulated Temperature Scanning Calorimetry (MTSC) experiment
for a PVAC sample which had been preliminary quickly cooled from a temperature
above its glass transition temperature. The MTSC results show a rise of the
specific heat which becomes smoother when the frequency of the temperature
modulation increases. This is expected and it is consistent with our
discussions of Sec.~\ref{sec:constantt}.
A calorimetry experiment with temperature modulated at frequency 
$\omega$ is only sensitive to degrees of freedom
which are faster that $2\pi / \omega$. For a system like PVAC, which has a
continuous spectrum of relaxation times, the variation of $\omega$ does not
detect qualitative changes when $\omega$ crosses specific relaxation
frequencies but instead a smoother change. This nevertheless shows that, when
the frequency of modulation increases, the number of channels which contribute
to the energy exchange at a given temperature decreases. As a result higher
temperatures are necessary to approach the equilibrium specific heat.

Although the DSC and MTSC experiments
have used the same heating rate for the temperature ramp, the DSC curve shows
a hump which is not observed in the MTSC measurements. One may ask weather this
is only a matter of time scales so that lower and lower values of $\omega$
could finally probe the hump observed in DSC, or if there is a fundamental
reason that prevents MTSC from detecting some of the phenomena that DSC
probes.
{\em This is an important question for calorimetry methodology. Using a test
  ``sample'' such as the three-state system allows us to give an unambiguous
  answer, as shown in the next section.}

\subsection{Analysis with a model system}
\label{sec:analysis}

As shown in the previous sections, the three state system can be used as a
sample system to bring further insights on the methods of calorimetry because
it allows a detailed analysis of the phenomena which is not possible from
experiments alone. The equivalent of a DSC experiment is obtained by imposing a
temperature ramp $T_1(t) = T_0 \pm s t$ where $s$ is a slope which measures the
variation of temperature per time unit and the $\pm$ sign allows for heating
or cooling. In a simulation we can record the energy $E(t)$ of the system and
compute its heat capacity $C_{DSC} = dE/dT$.

\begin{figure}[h]
  \centering
  \includegraphics[width=7cm]{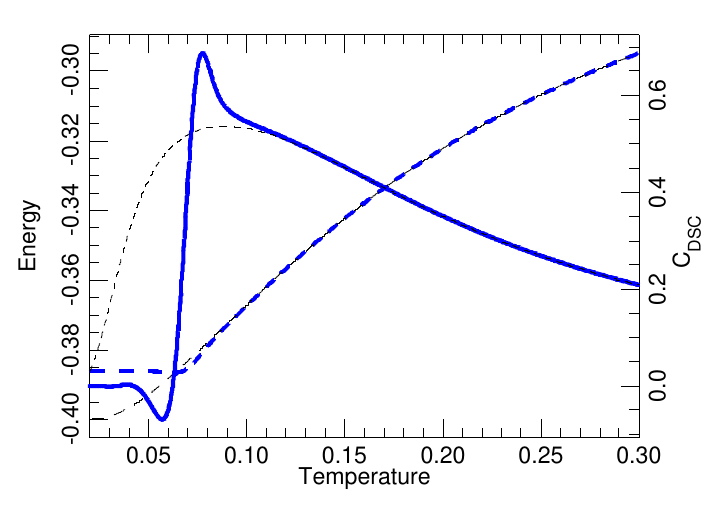}
  \caption{Results of the simulation of a DSC experiment with the three-state
    system. The thick dashed blue line shows the energy (left scale)
    versus temperature
    during a heating ramp from $T=0.02$ to $T=0.3$ in $10^{5}$t.u. after
    cooling from equilibrium at $T=0.3$ down to $T=0.02$ in the same time
    interval. The thin black
    dashed line shows the equilibrium energy
    $E^{\mathrm{eq}}(T)$ (Eq.~(\ref{eq:emoyeq})\,).
    The thick full blue line shows $C_{DSC}(T) = dE/dT$
    (right scale)
    during this heating scan and the thin black dotted line
    shows the equilibrium
    specific heat $C^{\mathrm{eq}}(T)$ (Eq.~(\ref{eq:ceq})\,). }
  \label{fig:dsc}
\end{figure}

Figure \ref{fig:dsc} shows the result of such a numerical experiment. The
three-state system, with the same parameters as above, was first cooled from
an equilibrium state at $T=0.3$ down to $T=0.02$ in $10^5\,$t.u.~. In the first
stage of the cooling its energy $E_{DSC}(t)$
followed the curve $E^{\mathrm{eq}}(T)$ but
below $T \approx 0.08$ it decreased slower and, at the end of the cooling scan
the system was out of equilibrium. Figure \ref{fig:dsc} shows the data for the
heating process from $T=0.02$ to $T=0.3$, which followed the cooling,
with a linear heating ramp in
$10^5\,$t.u.~. Below $T \approx 0.13$, the heat capacity $C_{DSC}(T)$ shows
strong deviations from the equilibrium heat capacity $C^{\mathrm{eq}}(T)$.
The sharp rise when temperature rises
  above $T \approx 0.07$, followed by a hump, occurs at
  the temperature at which the relaxation times of the system have
  sufficiently decreased to allow transitions between states during the
  characteristic time of the heating ramp. 
This is typical for a DSC scan with a sample initially out of equilibrium, and
displays
qualitative similarities with the hump observed experimentally with PVAC
(Fig.~\ref{fig:pvac}).

\medskip
The analogous of a MTSC experiment is obtained by adding a small temperature
modulation to the heating ramp
\begin{equation}
  \label{eq:ramp}
  T(t) = T_1(t) + \tac(t) = T_0 + s\, t + \dtac \, \sin \omega t \; .
\end{equation}
As a result, in addition to the increase of the energy due to the heating
ramp, the energy has an additional oscillatory component
$E_{\mathrm{ac}}(t)$. In an experiment this modulated part is usually
extracted by Fourier transform. In our simulations, it can be obtained by
subtracting $E_{DSC}(t)$ from the value $E_{MTSC}(t)$ recorded during the
simulation with the additional modulation $\tac(t)$.

\begin{figure}[h]
  \centering
  \includegraphics[width=7cm]{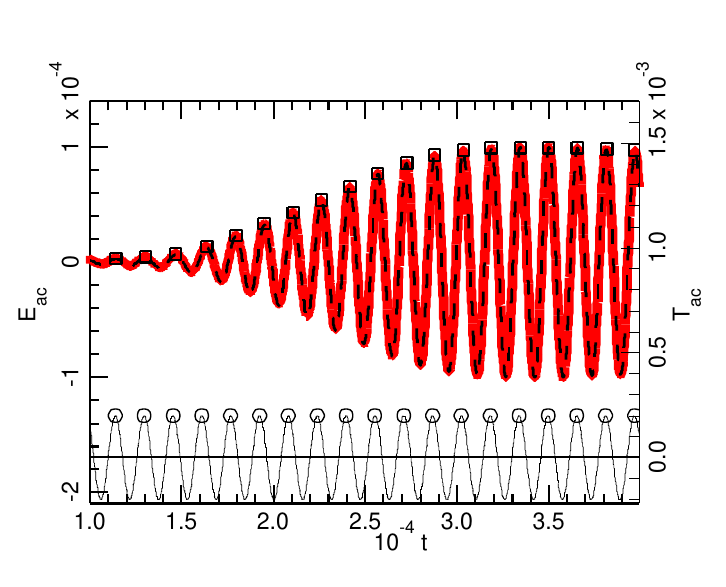}
  \caption{$E_{\mathrm{ac}}(t)$, (thick red line, left scale)
    the fraction of the energy which is modulated
    due to the modulated temperature $\tac$ in a simulation of an MTSC
    experiment ($\omega = 4\;10^{-3}\,$t.u.$^{-1}$).
    This figure only plots a fraction of the heating scan, which
    lasts from $t=0$ to $t=0.4 \; 10^5\,$t.u.
    in order to show the details of the
    modulation. The sinusoidal black line at the bottom shows $\tac(t)$ (right
    scale). The squares and circles mark the maxima of $E_{\mathrm{ac}}(t)$ and
    $\tac(t)$ which are used to calculate the modulus and phase of the
    specific heat as explained in the text. The dashed black line along the
    thick red line for $E_{\mathrm{ac}}(t)$ is an analytical result
    for $E_{\mathrm{ac}}(t)$ discussed in the text. }
  \label{fig:eac}
\end{figure}

Figure \ref{fig:eac} shows $E_{\mathrm{ac}}(t)$ for a portion of the heating
scan with $\dtac = 2\,10^{-4}$ and $\omega = 4\,10^{-3}\,$t.u.$^{-1}$
(corresponding to a period of $ \tau_{\omega} = 1571\,$t.u.).
The numerical results can be used
to calculate the magnitude $C(\omega)$ of the frequency-dependent heat
capacity and its phase $\Phi$,
using an approach which mimics the experimental approach. We look for
the values and dates of the maxima  $E_{\mathrm{ac}}^{\mathrm{max}}(t_j)$
of $E_{\mathrm{ac}}(t)$ and $T_{\mathrm{ac}}^{\mathrm{max}}(t_k)$ of
$\tac(t)$. Then for each maximum $k$ of $\tac(t)$, we look for the closest
maximum $j$ of $E_{\mathrm{ac}}(t)$ and we define
\begin{align}
  \label{eq:ampliphase}
  C(\omega,t_k) &=
  E_{\mathrm{ac}}^{\mathrm{max}}(t_j)/T_{\mathrm{ac}}^{\mathrm{max}}(t_k)
  \nonumber \\                
  \Phi(t_k) &= 2 \pi (t_j - t_k) /  \tau_{\omega}
  \nonumber \\               
\end{align}
The values of $C(\omega)$ and $\Phi(t_k)$ are obtained at time $t_k$,
when the temperature of the sample can be considered to be $T_1(t_k)$
because $\tac \ll T_1$.

\begin{figure}[h]
  \centering
  \includegraphics[width=7cm]{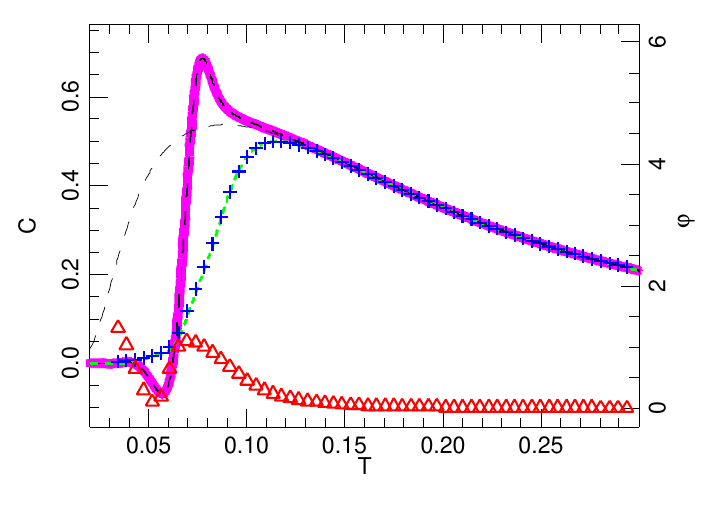}
  \caption{Heat capacity in a simulated MTSC experiment
    ($\omega = 4\;10^{-3}\,$t.u.$^{-1}$). The blue crosses show
    the magnitude of $C(\omega)$, determined with Eq.~(\ref{eq:ampliphase})
    from the maxima of $E_{\mathrm{ac}}$ and
    $\tac$,  versus the temperature $T_1(t)$ of the heating ramp. The
    red triangles
  show its phase $\Phi$ with respect to the temperature modulation. The figure
also shows $C_{DSC}(T)$ for the same sample system (thick magenta curve) and
the equilibrium heat capacity (thin black dashed line). The dash black
line which
follows  $C_{DSC}(T)$ and the dotted green line
along the $C(\omega)$ crosses are
analytical results discussed in the text.}
  \label{fig:cv1}
\end{figure}

Figure \ref{fig:cv1} shows the modulus of the dynamic heat capacity versus the
temperature $T_1(t)$ of the heating ramp, and its phase relative to the
temperature modulation. The equilibrium specific heat $C^{\mathrm{eq}}(T)$ and
$C_{DSC}(T)$ are also plotted for comparison. At low temperature $C(\omega)$
deviates significantly from  $C^{\mathrm{eq}}(T)$, which is expected because,
in this temperature range, thermal relaxation is very slow. The relaxation
times $\tau_1$ and $\tau_2$ for the two eigenmodes are well above the period
of the temperature modulation as shown in Fig.~\ref{fig:tau} so that the
response of the system to the modulation is weak. When temperature increases,
$\tau_1$ and $\tau_2$ decrease and $C(\omega)$ tends to
$C^{\mathrm{eq}}(T)$. For the same reason the phase shift $\Phi$ is large at low
temperature and tends to $0$ when $C(\omega)$ approaches
$C^{\mathrm{eq}}(T)$. Around $T=0.05$ the phase shift shows an oscillation
which could be related to contribution of the metastable states getting
destabilized by the temperature rise, but this is not reflected by any hump in
$C(\omega)$ which shows a monotonous rises towards $C^{\mathrm{eq}}(T)$, as
observed experimentally for PVAC (Fig.~\ref{fig:pvac}).

\begin{figure}[h]
  \centering
  \begin{tabular}{c}
    \textbf{(a)} \\
    \includegraphics[width=7cm]{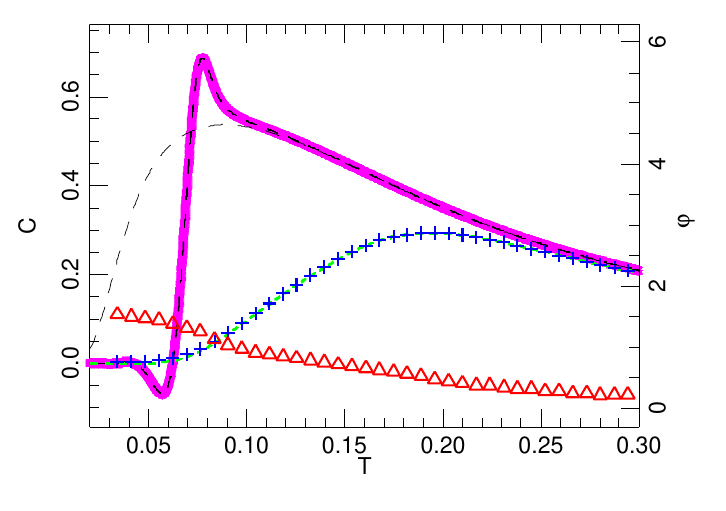} \\
    \textbf{(b)} \\
    \includegraphics[width=7cm]{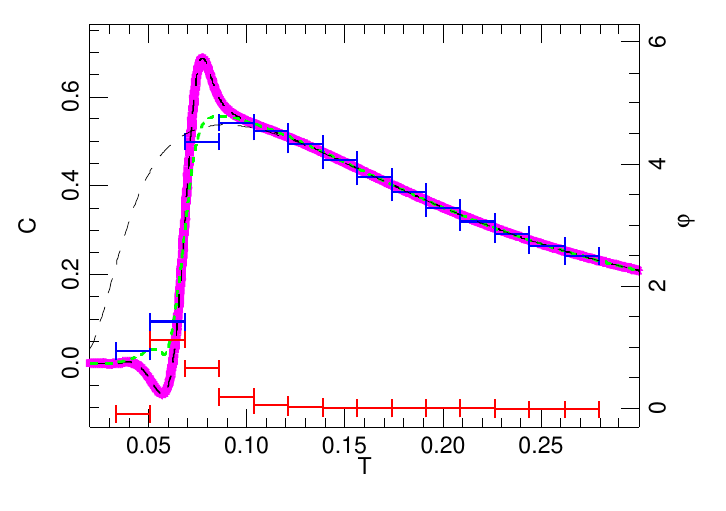}
  \end{tabular}
  \caption{Same as Fig.~\ref{fig:cv1} for two other values of the frequency of
    the temperature modulation:
    a) $\omega = 0.1\,$t.u.$^{-1}$ and b) $\omega = 10^{-3}\,$t.u.$^{-1}$.
  For this very slow modulation, the period  $\tau_{\omega} =
  0.63\,10^4\,$t.u. becomes significant with respect to the duration of the
  heating 
  ramp $10^5\,$t.u. and we have indicated the time interval between two maxima
  by horizontal error bars.}
  \label{fig:cv2}
\end{figure}

Figure \ref{fig:cv2} shows how these results depend on the frequency $\omega$
of the modulation. When $\omega$ increases (Fig.~\mbox{\ref{fig:cv2}-a})
the experiment probes faster and
faster time scales. Higher temperatures are needed to bring the relaxation
times of the system in the range of the time scales which are detected, so
that $C(\omega)$ grows more slowly with $T$ and only approaches
$C^{\mathrm{eq}}(T)$ at higher temperatures. Conversely, for very slow
modulations (Fig.~\mbox{\ref{fig:cv2}-b}), the maximum slope of $\tac(t)$
decreases 
so much that it is no longer much greater than the slope of the heating ramp.
In this case the rise of $C(\omega)$ versus $T$ becomes as fast as the
rise of $C_{DSC}(T)$ in the range $T \approx 0.05 - 0.08$. Nevertheless, in
agreement with the experimental observations (Fig.~\ref{fig:pvac}) the
simulation {\em does not show any anomaly such as the hump in specific heat
  observed in scanning calorimetry. This suggests that there is a fundamental
  difference between the results which can be obtained by DSC and by MTSC.}

\medskip
However numerical experiments, like actual experiments, provide {\em
  observations} but they do not give a full picture of the {\em mechanisms}
which act behind the scene to lead to these observations. Fortunately, working
with a tractable model system allows us to go beyond observations because
analytical calculations are possible to analyze the data and they are
revealing.

In Sec.~\ref{sec:constantt} and appendix
\ref{app:energyt} we showed that the time evolution of
the energy of the three-state system at temperature $T(t) = T_0 + \tac(t)$
can be calculated analytically when $T_0$ is a constant. The method can be
extended if $T_0$ is replaced by a temperature ramp $T_1(t)$, although the
solution cannot be expressed in closed form, by dividing the ramp in small
time intervals $\Delta t = [t_j , t_{j+1}]$. In such an interval $T_0$ is
replaced by $T_1(t_j)$ which can be treated as constant if $\Delta t$ is
sufficiently small with respect to the relaxation times $\tau_1[T_1(t_j)]$,
$\tau_2[T_1(t_j)]$. Knowing the occupation probabilities $P_1(t_j)$,
$P_2(t_j)$, the calculation presented in Appendix~\ref{app:energyt} allows us
to calculate  $P_1(t_{j+1})$, $P_2(t_{j+1})$ in the presence of the
temperature modulation. This defines the initial state for the next time
interval so that we can proceed step by step from the start to the end of the
temperature ramp. As the calculation proceeds the eigenmodes and
relaxation times $\tau_1$,
$\tau_2$ have to be recalculated according to the change of $T_1(t)$ from one
interval to the next. However, as this calculation is fast, we can select a very
small value for $\Delta t$ to ensure a good accuracy to the process which
would converge to an exact result in the limit $\Delta t \to 0$. In practice
we used $\tau_{\omega}/5000 \le \Delta t \le \tau_{\omega}/500$.

\medskip
The interest of this calculation is not to make sure that the analytical
calculation can reproduce the simulation results (which it does) but to
understand the origin of the observations. We showed that the analytical
calculation proceeds by expressing the occupation probabilities as
\begin{equation}
  \label{eq:PQq}
  P_i(t) = P_i(T_{\mathrm{ref}}) + \widetilde{Q_i}(t) + q_i(t) 
\end{equation}
$T_{\mathrm{ref}}$ being either $T_0$ or $T_1(t_j)$. 
$\widetilde{Q_i}(t)$ is the solution that we would get in the absence of
modulation and $q_i(t)$ is an additional contribution which is entirely due
to the modulation. As shown by Eq.~(\ref{eq:solutionEe}) the energy splits
into
\begin{equation}
  \label{eq:Ee}
  E(t) = \widetilde{E}(t) + e(t) \;
\end{equation}
where $\widetilde{E}(t)$ depends on $\widetilde{Q_i}(t)$ and $e(t)$ depends on
$q_i(t)$. This expression can therefore be directly related to the
experimental results. A temperature ramp without modulation corresponds to a
DSC experiment. Therefore $\widetilde{E}(t)$ should correspond to the energy
measured by DSC and $d \widetilde{E}(t) / dT_1(t)$ should give $C_{DSC}(T)$.
Figures \ref{fig:cv1} and \ref{fig:cv2} show that this is exactly the case:
the dashed black line which follows the magenta line deduced from the
numerical simulation of a DSC experiment (Fig.~\ref{fig:dsc}) plots
$d \widetilde{E}(t) / dT_1(t)$. Equation (\ref{eq:Ee}) also tells us that
$e(t) = E(t) - \widetilde{E}(t)$ is the contribution of the modulation to the
energy, i.e. $E_{\mathrm{ac}}(t)$ in an experiment. Again we can verify this
to a good accuracy on Fig.~\ref{fig:eac} as the dashed black line which
follows the thick red line plotting $E_{\mathrm{ac}}(t)$ in a simulated MTSC
experiment is the curve for $e(t)$ deduced from the analytical calculation.
Therefore the analytical derivative of $e(t)$ gives the dynamic heat
capacity of the system. We calculate the time derivative $de(t)/dt$ and
extract from this expression the prefactor of $\sin \omega t$ (in phase with
$T_{\mathrm{ac}}(t)$) and the prefactor of $\cos \omega t$ (in quadrature
of phase with $T_{\mathrm{ac}}(t)$). Dividing these two prefactors by the
amplitude of $d T_{\mathrm{ac}}(t)/dt = A_T \omega$ we get $C'(\omega)$ and
$C''(\omega)$. Then
the modulus of the dynamic heat capacity is readily obtained as
$C(\omega) = \big[ C'(\omega)^2 + C''(\omega)^2 \big]^{1/2}$. The green lines on
Figs.~\ref{fig:cv1} and
\ref{fig:cv2} show that the analytical expression of $C(\omega)$ exactly
follows the simulation results.

\bigskip
These results demonstrate that, for a glassy system out of equilibrium,
$C_{DSC}$ and $C(\omega)$ are fundamentally
different quantities. $C_{DSC}$ measures the relaxations due to the
temperature drift, i.e. the response to $T_1(t)$, while $C(\omega)$ only
selects the oscillatory response to $T_{\mathrm{ac}}(t)$. Therefore
{\em we should not expect that $C(\omega)$ should converge to $C_{DSC}$ in the
  limit $\omega \to 0$.} This explains why experimental measurements of these
two quantities differ as shown in Sec.~\ref{sec:experimental}.  $C(\omega)$ do
captures some components of $C_{DSC}$ because the slope of its rise when $T$
increases grows and tends towards the slope of the rise of $C_{DSC}(T)$, but
$C(\omega)$ misses the extra humps which are pure relaxation phenomena.
Actually, as shown in Sec.~\ref{sec:constantt}, relaxations are not entirely
absent from the dynamic heat capacity because they enter in a correction
factor for the amplitude of the response to $T_{\mathrm{ac}}(t)$
(see Eqs.~(\ref{eq:smallq}) and (\ref{eq:Gamma})~). Those corrections could
play a significant role for the temperature jumps discussed in
Sec.~\ref{sec:constantt}, but they become negligible when temperature varies
continuously. They might become noticeable if the variation $T_1(t)$ could
become very fast. But this cannot be the case in an scanning experiment which
intends to measure the response to an oscillatory temperature
component. Experimentally the oscillation can only be detected if the maximum
slope of $T_{\mathrm{ac}}(t)$ is larger than the slope of $T_1(t)$.
For $\omega = 10^{-3}\,$t.u.$^{-1}$ Fig.~\ref{fig:cv2} shows
that, even in this
extreme case with only about 16 periods of $T_{\mathrm{ac}}(t)$ during the
full heating ramp, the relaxation humps observed in $C_{DSC}(T)$ do not show
up in $C(\omega,T)$.

\section{Discussion}
\label{sec:discussion}

In this paper we have used a combination of numerical simulations and
analytical calculations for a model system, as well as comparisons with actual
experimental data for a glassy system, to provide a deeper understanding of
temperature modulated calorimetry measurements which have been at the origin
of many 
discussions \cite{GARDEN-REVIEW}. We have essentially considered two
questions:

\begin{itemize}
\item   how does the signal looks when one performs experiments with a 
small temperature modulation added to a constant underlying temperature ? In
particular we examined how the dynamic heat capacity may depend on time in
this case.

\item for scanning calorimetry with an underlying temperature
  which varies as a ramp, is there a fundamental difference between a
  standard calorimetry method such as DSC and a temperature modulated
  calorimetry  measurement such as MTSC ?

\end{itemize}

\medskip
In the case of a constant underlying temperature we showed that, as expected,
temperature modulated calorimetry probes the spectrum of the relaxation times
of the sample 
system. The energy oscillates with some phase shift with respect to the
temperature modulation and these oscillations are accompanied by an entropy
production which varies at twice the frequency of the temperature modulation
because entropy is created when temperature moves up and down.

As a sample
system we investigated the three-state system which is the simplest system
with a 
non-trivial, non-single-frequency spectrum of the fluctuations of
the energy transfers \cite{PG}.
This simple system allows a full analytical calculation of the
dynamic specific heat which has a component in phase with the temperature
modulation but also a component in quadrature with it. These two contributions
are often designated as a ``complex'' heat capacity. Using a complex notation
to compute $C(\omega)$ is however misleading because
it focuses on a steady response, and
does not explicitly introduces the boundary conditions in the solution.
Those boundary conditions may be important because
temperature modulated calorimetry measurements probe
not only the spectrum of the energy transfers in the sample but also the
time evolution of the state of the system, such as its aging. 
After a sharp temperature jump the system may undergo strong relaxations, and
we showed that these relaxations appear even in the oscillatory component of
the energy. In measurements they show up as a time-dependent heat capacity.
A spectacular example is provided by the Kovacs effect for glasses. To our
knowledge it has never been investigated by temperature modulated calorimetry,
although our 
study shows how such measurements could clarify its origin because the
duration of the Kovacs hump
is related to the time scales which govern the internal
evolution of the glass.

\medskip
In the case of scanning calorimetry experiments for
  out-of-equilibrium glassy systems, we showed
that standard experiments such as DSC and measurements with an oscillatory
temperature such as MTSC probe intrinsically different properties of the
system. The analytical
calculation explains why, even in the limit of very low modulation frequency,
some of the features observed in DSC do not show up in MTSC.
Besides the case of PVAC presented in Fig.~\ref{fig:pvac} and
  in ref.~\cite{LAARRAJ},
  the differences between DSC and MTSC experimental results
have been discussed in various papers
\cite{ANDROSCH,SARUYAMA,GARDEN2004,GARDEN2005}. All studies confirm that, at
high modulation frequency, the MTSC measurements detect a lower value of the
heat capacity than DSC because they are only sensitive to energy transfers
which are fast enough to take place within one period of the
modulation. However these studies show that it is important to distinguish
between experiments that study thermodynamic transitions, for instance in
paraffin \cite{SARUYAMA} or PTFE \cite{ANDROSCH,GARDEN2005}, and protein
folding \cite{GARDEN2004}, which could be observed at equilibrium or
quasi-equilibrium in very slow temperature scans,
from observations of out-of-equilibrium effects in glasses \cite{LAARRAJ}.
For equilibrium transitions, near the transition temperature, in the limit
$\omega \to 0$ the modulation of 
the temperature is sufficient to change the fraction of the sample which has
passed the transition. Therefore, in this case integrating
$C(\omega,T)$ versus $T$ around the transition temperature and taking the
limit $\omega \to 0$ recovers the value of the latent heat
\cite{GARDEN2005}. Conversely, experiments with glasses out of equilibrium, and
calculations for the three-state model, show that the relaxations detected by
$C_{DSC}$ are different from the dynamic heat capacity measured by
ac-calorimetry, even in the limit $\omega \to 0$. This is because the
evolution of the system is not caused by the temperature modulation.
Instead the strongly out-of-equilibrium initial state tends to spontaneously
evolve towards equilibrium when $T$ is raised in the DSC scan.

\medskip
Although the conclusions based on the analytical calculations have been
obtained with a particular model, the three-state system, we think that they
have a much broader validity because this model is very generic. It describes
a sample with a free energy landscape which has many metastable minima, and
which evolves between them under the effect of thermal fluctuations.
These are features which have been proposed for glasses and complex liquids
\cite{STILLINGER-82} as well as proteins \cite{NAKAGAWA}, but can be expected
to apply to many systems.
Numerical simulations of a bead-spring polymer model
known to be a glass-former have shown that, at low temperature, the low
frequency component of $C''(\omega)$ is entirely the result of
the dynamics of the system within
its inherent structures, i.e.\ the minima of its potential energy landscape
\cite{BROWN-JR2011}. With only two relaxation
times, the three-state system is the simplest example of a
large family which has a non-trivial spectrum. It is sufficient to test some
ideas on temperature modulated calorimetry, for 
instance by showing how a scan in the modulation frequency $\omega$ detects
one frequency after another. For a real system with many relaxation times, the
analytical calculations that we developed
in the appendices cannot be carried out in
practice, but the methods are, in principle, still valid, at the expense of
large matrices of eigenstates. Therefore, the qualitative aspects of
our results should be preserved, such as for instance the fundamental
difference 
between DSC and MTSC results for a glass.
This assumption is supported by the results on
PVAC that we presented. And actually, as shown in \cite{PG} the three-state
system itself could be of interest to analyze various experimental
observations, in cases where the simple two-state system, with a single
relaxation, fails.
This model only describes the configurational heat
capacity. This is indeed a limitation but, in many systems it is however the
contribution which is the most interesting because vibrational or electronic
contributions are generally smoother versus temperature or significantly
weaker \cite{TOMBARI2007}.

\begin{acknowledgments}
  The authors would like to thank M. Laarraj who performed the TMSC
  experiments on PVAC used in Figs.~\ref{fig:ClogtPVAC}
  and \ref{fig:pvac}, and J. Richard
  (Institut Néel, Grenoble, France) for useful discussions
  and for the accurate treatment of experimental data 
which lead to Fig.~\ref{fig:pvac}.
\end{acknowledgments}

\appendix
\section{Analytical calculation of the energy of the three-state system
  at modulated temperature.}
\label{app:energyt}

We consider the case of a temperature modulation around a fixed average
value $T_0$, $T(t) = T_0 + \tac(t) = T_0 + A_T \sin \omega t$.
To get the energy $E(t) = \sum_{i=1}^3 P_i(t) E_i$, given an initial state of
the system at time $t=0$ determined by $P_i(t=0)$, we must solve the set of
equations (\ref{eq:dp1dt}) and the similar equations for $P_2$ and $P_3$,
with $\omega_{ij} = 1$ and the condition $\sum_{i=1}^3 P_i = 1$.

It is convenient to introduce the variables $Q_i(t) = P_i(t) -
P_i^{\mathrm{eq}}(T_0)$ where $P_i^{\mathrm{eq}}(T_0)$,
henceforth denoted by $P_i^0$, are the equilibrium
probabilities at temperature $T_0$. The condition $\sum_{i=1}^3 P_i = 1$
implies $\sum_{i=1}^3 Q_i = 0$, so that $E(t)$ is determined by $Q_1$ and $Q_2$
only. Note that {\em we make no assumption regarding the size of $Q_i$
  compared to $P_i^{\mathrm{eq}}$. In strongly out-of-equilibrium situations it
  may happen that $|Q_i/P_i^{\mathrm{eq}}| \gg 1$.}

\smallskip
As we assume $\tac \ll T_0$, the rates of the thermally activated
transitions $W_{ij}(t)$ (Eq.~\ref{eq:wij}) can be
expanded around their values $W_{ij}^0$ at temperature $T_0$  to
first order in $\tac$ as
\begin{equation}
  \label{eq:expanw}
  W_{ij}(t) = W_{ij}^0 + \tac(t) \frac{B_{ij}}{T_0^2} W_{ij}^0 \; .
\end{equation}
Using these expansions, the equation for $dQ_1/dt$, deduced from
Eq.~(\ref{eq:dp1dt}) splits into 4 components
\begin{equation}
  \label{eq:dq1components}
  \frac{dQ_1}{dt} = C_1 + C_2 + C_3 + C_4 \;,
\end{equation}
with
\begin{align}
  \label{eq:defci}
  C_1 &= - P_1^0 W_{12}^0 + P_2^0 W_{21}^0 - P_1^0 W_{13}^0 + P_3^0 W_{31}^0
        \nonumber \\
  C_2 &= -Q_1 \left[ W_{12}^0 + W_{13}^0 + W_{31}^0 \right]
        + Q_2 \left[ W_{21}^0 - W_{31}^0 \right]
        \nonumber \\
  C_3 &= \tac \frac{1}{T_0^2} \left[ -P_1^0 B_{12} W_{12}^0 - P_1^0 B_{13}
        W_{13}^0 \right. \nonumber \\
      &\qquad \qquad\left. + P_2^0 B_{21} W_{21}^0  + P_3^0 B_{31} W_{31}^0
        \right]  \nonumber \\
  C_4 &= - Q_1 \frac{\tac}{T_0^2} \left[ B_{12} W_{12}^0 + B_{13} W_{13}^0
        + B_{31} W_{31}^0 \right] \nonumber \\
      &~ \qquad + Q_2 \frac{\tac}{T_0^2} \left[ B_{21} W_{21}^0 - B_{31} W_{31}^0
        \right] \nonumber \\ \; .
\end{align}
Component $C_1$ vanishes due to the detailed balance condition at temperature
$T_0$. Component $C_2$, in which
we used $Q_3 = - (Q_1 + Q_2)$,
is of the form $C_2 = -A Q_1 + B Q_2$ if we introduce
the notations $A$ and $B$ of the two brackets that it contains. Component
$C_3$ can be written $C_3 = \tac \gamma_1$ by introducing a notation for the
bracket divided by $T_0^2$ and component $C_4$ can be written
$C_4 = - C Q_1 \tac + D Q_2 \tac$ by introducing $C$ and $D$ to designates the
two brackets divided by $T_0^2$. The quantities $A$, $B$, $C$, $D$ and
$\gamma_1$ are time independent, while $\tac = \dtac \sin \omega t$
depends on time. A similar calculation for the time evolution of $Q_2$ leads
to
\begin{equation}
  \label{eq:q2}
  \frac{dQ_2}{dt} = B' Q_1 - A' Q_2 + \tac \gamma_2 + D' Q_1 \tac - C' Q_2
  \tac
\end{equation}
with
\begin{align}
  A' &= W_{21}^0 + W_{23}^0 + W_{32}^0 \nonumber\\
  B' &= W_{12}^0 - W_{32}^0 \nonumber\\
  \gamma_2 &= \frac{1}{T_0^2} \left[ -P_2^0 B_{21} W_{21}^0 - P_2^0 B_{23}
             W_{23}^0 \right. \nonumber \\
  & \qquad \qquad \left. + P_1^0 B_{12} W_{12}^0 + P_3^0 B_{32} W_{32}^0 \right]
  \nonumber \\
  C' &= \frac{1}{T_0^2} \left[ B_{21} W_{21}^0 + B_{23} W_{23}^0 + B_{32}
  W_{32}^0 \right] \nonumber \\
  D' &= \frac{1}{T_0^2} \left[B_{12} W_{12}^0 - B_{32} W_{32}^0 \right]
 \nonumber \\
\end{align}
The equations for $Q_1$ and $Q_2$ can therefore be put in the matrix form
\begin{align}
  \label{eq:qmatrix}
  \frac{d}{dt}
  \left( \begin{array}{c}Q_1 \\ Q_2 \end{array} \right)
  &=
    \left( \begin{array}{cc} -A & B \\ B' & -A' \end{array} \right)
   \left( \begin{array}{c}Q_1 \\ Q_2 \end{array} \right)
 \nonumber \\ &~ + \tac \left[ 
 \left( \begin{array}{c}\gamma_1 \\ \gamma_2 \end{array} \right)
 + \left( \begin{array}{cc} -C & D \\ D' & -C' \end{array} \right)
 \left( \begin{array}{c}Q_1 \\ Q_2 \end{array} \right) \right]
\nonumber \\
  &= M  \left( \begin{array}{c}Q_1 \\ Q_2 \end{array} \right) +
\tac \left[ \left( \begin{array}{c}\gamma_1 \\ \gamma_2 \end{array} \right)
+ N \left( \begin{array}{c}Q_1 \\ Q_2 \end{array} \right) \right] \;,
\end{align}
where we have introduced two matrices $M$ and $N$.
The first term in the right hand side determines the solution
in the absence of temperature modulation that we denote by $\widetilde{Q_1}$,
$\widetilde{Q_2}$, which was studied in \cite{PG}.
To solve
\begin{align}
  \label{eq:systemeqtilde}
  \frac{d}{dt}  \left( \begin{array}{c}\widetilde{Q_1} \\
                         \widetilde{Q_2} \end{array} \right)
  = M  \left( \begin{array}{c}\widetilde{Q_1} \\ \widetilde{Q_2} \end{array}
  \right)  
\end{align}
we can expand $\widetilde{Q_1}$ $\widetilde{Q_2}$ on the eigenvectors
$\vec{U}^{(1)}$ 
and $\vec{U}^{(2)}$ which diagonalize 
the matrix $M$
\begin{align}
  \label{eq:defvectpr}
  M \vec{U}^{(i)} = \lambda_i \vec{U}^{(i)} \; .
\end{align}
The eigenvalues  $\lambda_1$ and $\lambda_2$ are
\begin{equation}
  \label{eq:lambda}
  \lambda_{1,2} = \frac{1}{2} \left[ -(A + A') \pm \sqrt{\Delta} \right]
\end{equation}
with $\Delta = (A - A')^2 + 4 B B'$.
The system parameters which are compatible with the existence of a thermal
equilibrium are such that $\lambda_{1,2} < 0$. Each eigenvalue corresponds to
an eigenvector $\vec{U}^{(i)}$ ($i = 1,2$). Its components are denoted as
\begin{align}
  \label{eq:composantesvectpr}
 \vec{U}^{(i)} = \left( \begin{array}{c} U_1^{(i)} \\ U_2^{(i)} 
                      \end{array} \right)
\end{align}

Matrix $M$ is not a symmetric matrix. It is not orthogonal and it is easy to
check that its eigenvectors are not orthogonal to each other, i.e.
\begin{equation}
  \label{eq:nonortho}
  U_1^{(1)} \; U_1^{(2)} + U_2^{(1)} \; U_2^{(2)} \not= 0 \; .
\end{equation}
However those vector are not colinear
\begin{equation}
  \label{eq:noncolin}
  U_1^{(1)} \; U_2^{(2)} - U_2^{(1)} \; U_1^{(2)} \not= 0
\end{equation}
and therefore they nevertheless define a basis for the $\widetilde{Q_1}$,
$\widetilde{Q_2}$ space.  On
this basis $\vec{\widetilde{Q}}$ can be written as
  \begin{equation}
    \label{eq:decomposQ-app}
    \vec{\widetilde{Q}} = a(t) \vec{U}^{(1)} + b(t) \vec{U}^{(2)} \; .
  \end{equation}
Equation  (\ref{eq:decomposQ-app})  defines a system of two scalar equations for
$a$ and $b$. Its determinant is
\begin{align}
  \label{eq:determinant}
  D_U = \left| \begin{array}{cc} U_1^{(1)} & U_1^{(2)} \\
                               U_2^{(1)} & U_2^{(2)}
                      \end{array} \right| \; .
\end{align}
It does not vanish due to the relation (\ref{eq:noncolin}).
Solving 
Eq.~(\ref{eq:systemeqtilde}) leads to
\begin{align}
  \label{eq:systemab}
  \frac {da(t)}{dt} \vec{U}^{(1)} + \frac{db(t)}{dt} \vec{U}^{(2)}
  = \lambda_1 a(t) \vec{U}^{(1)} + \lambda_2 b(t) \vec{U}^{(2)} \;
\end{align}
which can be viewed as a system of two equations for the unknowns
\begin{equation}
  \label{eq:defXY}
  X = \frac {da(t)}{dt} - \lambda_1 a(t)
  \qquad Y = \frac {db(t)}{dt} - \lambda_2 b(t) \; ,
\end{equation}
which can be written
\begin{align}
  \label{eq:systemeXY}
  U_1^{(1)} X + U_1^{(2)} Y &= 0 \nonumber \\
  U_2^{(1)} X + U_2^{(2)} Y &= 0  \; .
\end{align}
The determinant of this system is again the determinant $D_U$ of
Eq.~(\ref{eq:determinant}), which is non-zero. As the right-hand-side of the
system is zero, the only solution of the system is $X=0$, $Y=0$. According to
(\ref{eq:defXY}) it implies that the general solutions for $a(t)$ and $b(t)$
are exponential relaxations
\begin{align}
  \label{eq:sola-app}
  a(t) &= a(t_0) \exp[ - (t-t_0) / \tau_1] \\
  \label{eq:solb-app}
  b(t) &= b(t_0) \exp[ - (t-t_0) / \tau_2]
\end{align}
where
$\tau_{1,2} = -1/\lambda_{1,2}$. The values of $a(t=0)$ and $b(t=0)$ are
determined by the initial state of the system, which are assumed to be known
so that $\widetilde{Q_1}$ $\widetilde{Q_2}$ are fully determined by
Eqs.~(\ref{eq:decomposQ-app}) and (\ref{eq:sola-app}), (\ref{eq:solb-app}).

This solution for $\widetilde{Q_1}$ $\widetilde{Q_2}$, which corresponds to the
evolution of the system in the absence of the modulation $\tac$ of the
temperature, exhibits two relaxation times $\tau_1$, $\tau_2$. They
make up the spectrum of the thermal relaxations of the three-state system
which determines how an out-of-equilibrium state relaxes but also the response
to the temperature modulation $\tac$.

\medskip
Let us denote by $q_1$, $q_2$ this response to $\tac$, i.e. look for a
solution of Eq.~(\ref{eq:qmatrix}) under the form $Q_1 = \widetilde{Q_1} +
q_1$ and 
$Q_2 = \widetilde{Q_2} + q_2$. They are solution of
\begin{equation}
  \label{eq:smallq}
  \frac{d}{dt} \left( \begin{array}{c}q_1 \\ q_2 \end{array} \right)
  = M  \left( \begin{array}{c}q_1 \\ q_2 \end{array} \right)
  + \tac \left[ \left( \begin{array}{c}\gamma_1 \\ \gamma_2 \end{array} \right)
+ N \left( \begin{array}{c}Q_1 \\ Q_2 \end{array} \right) \right]
\end{equation}
Since we assumed that $\tac \ll T_0$ the response to the modulation
is itself of secondary order, so that, in the last term of
Eq.~\ref{eq:smallq}, which is of the order of $\tac$ we can replace
$Q_1$, $Q_2$ by $\widetilde{Q_1}$ $\widetilde{Q_2}$, which have been obtained
above. 

Defining
\begin{align}
  \label{eq:Gamma}
  \Gamma_1 &= \gamma_1 - C \widetilde{Q_1} + D \widetilde{Q_2}
             \nonumber \\
  \Gamma_2 &= \gamma_2 + D' \widetilde{Q_1} - C' \widetilde{Q_2}
             \nonumber \\
\end{align}
the equation for $q_1$, $q_2$ becomes
\begin{equation}
  \label{eq:smallqmatrix}
  \frac{d}{dt} \left( \begin{array}{c}q_1 \\ q_2 \end{array} \right)
  = M  \left( \begin{array}{c}q_1 \\ q_2 \end{array} \right) +
  \tac \left( \begin{array}{c}\Gamma_1 \\ \Gamma_2 \end{array} \right) \; .
\end{equation}
The last term of its right hand side is fully known once $\widetilde{Q_1}$
and $\widetilde{Q_2}$ have been computed. Expanding $q_1$ and $q_2$
on the eigenstates of matrix $M$ as
\begin{equation}
    \label{eq:decompos-smallq}
    \vec{q} = \alpha(t) \vec{U}^{(1)} + \beta(t) \vec{U}^{(2)} \; .
\end{equation}
the calculation of $\alpha(t)$ and $\beta(t)$ to get the
response to $\tac$ can proceed along the same lines as the derivation of
$a(t)$ and $b(t)$ presented above. It amounts to solving
\begin{align}
  \label{eq:alpha}
  \frac{d \alpha}{dt} + \alpha/\tau_1 &= \tac \delta_1 \\
  \label{eq:beta}
   \frac{d \beta}{dt} + \beta/\tau_2 &= \tac \delta_2 
\end{align}
with
\begin{equation}
  \label{eq:delta12}
  \delta_1 = \frac{1}{D_U} \left|
    \begin{array}{cc} \Gamma_1 & U_1^{(2)} \\ \Gamma_2 & U_2^{(2)}
    \end{array} \right|
  \qquad \delta_2 = \frac{1}{D_U} \left|
    \begin{array}{cc} U_1^{(1)} & \Gamma_1 \\ U_2^{(1)} & \Gamma_2
        \end{array} \right| 
\end{equation}

With $\tac = A_T \sin \omega t$ the solution of Eq.~\ref{eq:alpha} is obtained
by solving this equation without the right-hand-side to get $\alpha(t) =
\alpha_0 \exp[-(t-t_0)/\tau_1]$ and then plug this expression
into the full equation
assuming that $\alpha_0$ depends on time. This gives an equation for $d
\alpha_0/dt$ which can be integrated to give
\begin{align}
  \label{eq:solalpha1}
  \alpha(t) = & \alpha(t_0) e^{-(t-t_0)/\tau_1} + \frac{A_T
                \delta_1}{\sqrt{\omega^2 + 1/\tau_1^2}} \times
                \nonumber \\
  &\Big[ \sin(\omega t + \phi_1)
  - e^{-(t-t_0)/\tau_1} \sin(\omega t_0 + \phi_1) \Big] \,
\end{align}
with
\begin{equation}
  \label{eq:phi1}
  \tan \phi_1 = - \omega \tau_1 \; .
\end{equation}
The solution for $\beta$ in
Eq.~(\ref{eq:beta}) is similar with $\tau_2$, $\delta_2$,
$\phi_2$.

In the particular case $t_0 = 0$ we get
\begin{align}
  \label{eq:solalphabeta}
  \alpha &= \frac{ \delta_1 A_T}{\sqrt{\omega^2 + (1/\tau_1)^2}}
  \big[ \sin (\omega t + \phi_1) - \sin \phi_1 \big]
  \nonumber \\
  \beta &= \frac{ \delta_2 A_T}{\sqrt{\omega^2 + (1/\tau_2)^2}} 
  \big[ \sin (\omega t + \phi_2) - \sin \phi_2 \big]
          \; .           
\end{align}
Summarizing we get the solution for $Q_1$ $Q_2$ as
\begin{equation}
  \label{eq:sol1Q12}
  \left( \begin{array}{c} Q_1 \\ Q_2 \end{array} \right)  =
  \left( \begin{array}{c} \widetilde{Q_1} \\ \widetilde{Q_2} \end{array}
  \right)  + 
  \left( \begin{array}{c} q_1 \\ q_2 \end{array} \right)
\end{equation}
with
\begin{align}
  \label{eq:solutionQtilde}
\left( \begin{array}{c} \widetilde{Q_1} \\ \widetilde{Q_2} \end{array} \right)
  & = 
a(t=0) e^{-t / \tau_1}  \left( \begin{array}{c} U_1^{(1)} \\
                                 U_2^{(1)} \end{array} \right) 
\nonumber \\
  & + b(t=0) e^{-t / \tau_2}  \left( \begin{array}{c} U_1^{(2)} \\
                                   U_2^{(2)} \end{array} \right)
\end{align}
and
\begin{align}
  \label{eq:solutionqsmall}
  \left( \begin{array}{c} q_1 \\ q_2 \end{array} \right) &=
  \frac{ \delta_1 A_T}{\sqrt{\omega^2 + (1/\tau_1)^2}}
  \sin (\omega t + \phi_1)
\left( \begin{array}{c} U_1^{(1)} \\
                                 U_2^{(1)} \end{array} \right)
\nonumber \\
  & + \frac{ \delta_2 A_T}{\sqrt{\omega^2 + (1/\tau_2)^2}}
  \sin (\omega t + \phi_2)
 \left( \begin{array}{c} U_1^{(2)} \\
                                   U_2^{(2)} \end{array} \right)
\end{align}
The energy $E(t)$ is finally given by
\begin{align}
  \label{eq:solutionEt}
  E(t) &= (P_1^{\mathrm{eq}}(T_0) + Q_1)(E_1 - E_3) \nonumber \\ & \quad +
  (P_2^{\mathrm{eq}}(T_0) + Q_2)(E_2 - E_3) + E_3 \; ,
\end{align}
i.e.
\begin{align}
  \label{eq:solutionEe}
  E(t) &= (P_1^{\mathrm{eq}}(T_0) + \widetilde{Q_1})(E_1 - E_3)
         \nonumber \\ & \quad
      +(P_2^{\mathrm{eq}}(T_0) + \widetilde{Q_2})(E_2 - E_3) + E_3 
  \nonumber \\                                                                 
  & \quad + q_1 (E_1 - E_3) + q_2 (E_2 - E_3)
    \nonumber \\
       &= \widetilde{E(t)} + e(t)
\end{align}
where the last term $e(t)$ designates the contribution which is due
to the temperature modulation, while $\widetilde{E}$ is the contribution due
to the relaxation from the initial state if it was not already at equilibrium
at temperature $T_0$.
\section{Modulation-dependent heat capacity}
\label{app:heatcapacity}

This appendix again considers the case $T(t) = T_0 + \tac = T_0 + A_T \sin
\omega t$.
The heat capacity is given by $C(T_0,t) = (dE/dt) / (d \tac / dt)$.
As shown in Appendix \ref{app:energyt}, the energy can be split in two parts,
$\widetilde{E}$ which does not depend on the temperature modulation, and a
contribution which would not exist without the modulation. Let us henceforth
denote by $C_{\omega}(t)$ the specific heat which is associated to the
modulation. This is this contribution which is measured by temperature modulated calorimetry
\begin{equation}
  \label{eq:comeg}
  C_{\omega}(t) = \frac{d e(t)/dt}{(d \tac / dt)} = \frac{1}{\omega A_T \cos
    \omega t} \; \frac{d e(t)}{dt}
\end{equation}
The calculation of ${d e(t)/dt}$ is straightforward from the expression of
$e(t)$ given in Appendix \ref{app:energyt}, but in doing this derivation,
{\em one
should not forget that $\delta_1$ and $\delta_2$  may depend on time if
  the initial state of a measurement was not an equilibrium state at
  temperature $T_0$} because they depend on $\Gamma_1$, $\Gamma_2$
given by Eq.~(\ref{eq:Gamma}) which are functions of $\widetilde{Q_1}$ and
$\widetilde{Q_2}$.

\smallskip
Expanding the trigonometric functions which show up in the results,
such as $\sin (\omega t + \phi_1)$, in terms of $\cos \omega t$
and $\sin \omega t$, we can distinguish in
$d e(t) / dt$ the contribution which is in phase with $d \tac / dt$
and a contribution with a phase lag of $\pi / 2$ with
$d \tac / dt$. The expression of  $d e(t) / dt$ can be written as
\begin{equation}
  \label{eq:ecossin}
  \frac{d e(t)}{dt} = \Delta e_1 \, \cos \omega t
  + \Delta e_2 \, \sin \omega t 
\end{equation}

This allows us to define
\begin{equation}
  \label{eq:cprimesec}
  C'_{\omega}(t) = \frac{ |\Delta e_1 |}{\omega A_T} \qquad
  C''_{\omega}(t) = \frac{ |\Delta e_2 |}{\omega A_T} \; .
\end{equation}
These two terms, in phase with the temperature modulation and
in quadrature with it, 
correspond to the real and imaginary part of the modulation-dependent
specific heat, when a complex notation is used.


\begin{thebibliography}{xx}

\bibitem{KRAFTMAKHER}
Y. Kraftmakher,
{\it Modulated calorimetry and related techniques,}
Physics Reports {\bf 356}, 1-117 (2002)

\bibitem{GMELIN}
E. Gmelin,
{\it Classica1 temperature-modulated calorimetry: A review,}
Thermochimica Acta {\bf 304-305}, 1-26 (1997)
  
\bibitem{HATTA}
I. Hatta and A.J. Ikushima,
{\it Studies on Phase Transitions by AC Calorimetry,}
Jpn. J. Appl. Phys. {\bf 20}, 1995-2011 (1981)

\bibitem{MENCZEL}
J.D. Menczel and L. Judovits,
{\it Preface of a special issue of the Journal of Thermal Analysis on
  Temperature-Modulated Differential Scanning Calorimetry}
J. Thermal Analysis {\bf 54} 409-410 (1998)

\bibitem{GARDEN-REVIEW}
J.-L. Garden,
{\it Macroscopic non-equilibrium thermodynamics in dynamic calorimetry,}
Thermochimica Acta {\bf 452}, 85-105 (2007)

\bibitem{SCHAWE}
J.E.K. Schawe,
{\it A comparison of different evaluation methods in modulated temperature
  DSC,}
Thermochimica Acta {\bf 260}, 1-16 (1995)

\bibitem{READING1997}
M. Reading,
{\it Comments on "A comparison of different evaluation methods
  in modulated-temperature DSC,}
Thermochimica Acta {\bf 292}, 179-187 (1997)

\bibitem{ANDROSCH}
R. Androsch,
{\it Reversibility of the Low-Temperature Transitions of
Polytetrafluoroethylene as Revealed by Temperature- Modulated Differential
Scanning Calorimetry,}
J Polym Sci B: Polym Phys {\bf 39}, 750-756 (2001)

\bibitem{TOMBARI2007}
E. Tombari, C. Ziparo, G. Salvetti and G. P. Johari,
{\it Vibrational and configurational heat capacity of poly(vinyl acetate)
  from dynamic measurements,}
J. Chem. Phys. {\bf 127}, 014905-1-6 (2007)

\bibitem{NIELSEN}
J.K. Nielsen and J.C. Dyre,
{\it Fluctuation-dissipation theorem for frequency-dependent specific heat,}
Phy. Rev. B {\bf 54}, 15754-1-8 (1996)

\bibitem{BROWN-JR2009}
J.R. Brown, J.D. McCoy, and D.B. Adolf,
{\it Driven simulations of the dynamic heat capacity,}
J. Chem. Phys. {\bf 131}, 104507-1-5 (2009)

\bibitem{BROWN-JR2011}
J.R. Brown and J.D. McCoy,
{\it The potential energy landscape contribution to the dynamic heat capacity,}
J. Chem. Phys. {\bf 134}, 194503-1-6 (2011)

\bibitem{PG}
M. Peyrard and J.-L. Garden,
{\it Memory effects in glasses: Insights into the thermodynamics of
  out-of-equilibrium systems revealed by a simple model of the Kovacs effect,}
Phys. Rev. E {\bf 102}, 052122-1-13 (2020)

\bibitem{JOHARI1999}
G.P Johari, C. Ferrari, E. Tombari and G. Salvetti,
{\it Temperature modulation effects on a material’s properties:
  Thermodynamics and dielectric relaxation during polymerization,}
J. Chem. Phys. {\bf 110}, 11592-11598 (1999)

\bibitem{LAARRAJ}
M. Laarraj, R. Adhiri, S. Ouaskit, M. Moussetad,
C. Guttin, J. Richard, and J.-L. Garden,
{\it Highly sensitive pseudo-differential ac-nanocalorimeter for the study
  of the glass transition,}
Rev. Sci. Instrum. {\bf 86}, 115110-1-13 (2015)

\bibitem{STILLINGER-82}
F.H. Stillinger and T.A. Weber,
{\it Hidden structure in liquids,}
Phys. Rev. A {\bf 25}, 978-989 (1982)

\bibitem{BISQUERT}
J. Bisquert,
{\it Master equation approach to the non-equilibrium negative specific heat
at the glass transition,}
Am. J. Phys. {\bf 73}, 735-741 (2005)

\bibitem{DEBOLT}
M.A. DeBolt, A.J. Easteal, P.B. Macedo and C.T. Moynihan,
{\em Amalysis of Structural Relaxation in Glass Using Rate Heating Data}
J. Am. Ceram. Soc. {\bf 59}, 16-21 (1976)

\bibitem{WANG-JOHARI}
J. Wang and G.P. Johari,
{\it Effects of sinusoidal temperature and pressure modulation on
the structural relaxation of amorphous solids,}
J. Non-Cryst Solids {\bf 281}, 91-107 (2001)

\bibitem{THOMAS1931}
S.B. Thomas and G.S. Park,
{\it Studies on glass: IV. Some Specific Heat Data on Boron Trioxide,}
J. Phys. Chem. {\bf 35}, 2091-2102 (1931)

\bibitem{FIORE}
C.E. Fiore and M.J. de Oliveira,
{\it Entropy production and heat capacity of systems under time-dependent
  oscillating temperature,}
Phys. Rev. E {\bf 99}, 052131-1-7 (2019)

\bibitem{TOMBARI2002}
E. Tombari, S. Presto, G. Salvetti, and G. P. Johari,
{\it Spontaneous decrease in the heat capacity of a glass,}
J. Chem. Phys. {\bf 117}, 8436-8441 (2002)

\bibitem{GARDEN-RICHARD}
J.-L. Garden and J. Richard,
{\it Entropy production in ac-calorimetry,}
Thermochimica Acta {\bf 461}, 57-66 (2007)

\bibitem{KOVACS}
A.J. Kovacs,
{\it Transition vitreuse dans les polymères amorphes. Etude
  phénoménologique,}
Fortsch. Hochpoly.-Forsch., {\bf 3}, 394-507 (1963)

\bibitem{SARUYAMA}
Y. Saruyama,
{\it AC calorimetry at the first order transition point,}
J. Therm. Anal. {\bf 38}, 1827-1833 (1992)

\bibitem{GARDEN2004}
J.-L. Garden, E. Ch\^{a}teau, and J. Chaussy,
{\it Highly sensitive ac nanocalorimeter for microliter-scale liquids
  or biological samples,}
Appl. Phys. Lett. {\bf 84}, 3597-3599 (2004)

\bibitem{GARDEN2005}
E. Ch\^{a}teau, J.-L. Garden, O. Bourgeois, and J. Chaussy,
{\it Physical kinetics and thermodynamics of phase transitions probed
  by dynamic nanocalorimetry,}
Appl. Phys. Lett. {\bf 86}, 151913-1-3 (2005)

\bibitem{NAKAGAWA}
N. Nakagawa and M. Peyrard,
{\it The Inherent Structure Landscape of a Protein,}
 Proc. Natl. Acad. Sci. USA (PNAS) {\bf 103}, 5279-5284 (2006)

\end{thebibliography}
\end{document}